\documentclass[preprintnumbers,
amsmath,amssymb,floatfix,10pt,prd,twocolumn,
superscriptaddress,longbibliography,nofootinbib]{revtex4-2}
\usepackage{bm}
\usepackage{amsfonts}
\usepackage{latexsym}
\usepackage{amsmath}
\usepackage{palatino}
\usepackage{mathpazo}
\usepackage{textcomp}
\usepackage{float}
\usepackage{booktabs}
\usepackage{dcolumn}
\usepackage{dsfont}
\usepackage{ragged2e}
\usepackage{epsfig}
\usepackage[dvipsnames]{xcolor}
\usepackage{hyperref}
\hypersetup{           
    colorlinks=true,                
    breaklinks=true,                
    urlcolor= OrangeRed,                
    linkcolor= magenta,                
    bookmarksopen=false,
    filecolor=black,
    citecolor=red,
    linkbordercolor=blue}
\usepackage{graphicx}
\usepackage{orcidlink}
\usepackage[T1]{fontenc} % For better handling of hyphenation and accented characters

\begin{document}
\title{Dynamics of Geodesics in Non-linear Electrodynamics Corrected Black Hole and Shadows of its Rotating Analogue}

\author{Hari Prasad Saikia \orcidlink{0009-0008-9048-7719}}%
 \email{hariprasadsaikia@dibru.ac.in}
\affiliation{%
 Department of Physics, Dibrugarh University, Dibrugarh \\
 Assam, India, 786004}

\author{Mrinnoy M. Gohain\orcidlink{0000-0002-1097-2124}}
\email{mrinmoygohain19@gmail.com}
\affiliation{%
 Department of Physics, Dibrugarh University, Dibrugarh \\
 Assam, India, 786004}
\affiliation{%
 Department of Physics, DHSK College, Dibrugarh \\
 Assam, India, 786001}
 
\author{Kalyan Bhuyan\orcidlink{0000-0002-8896-7691}}%
 \email{kalyanbhuyan@dibru.ac.in}
\affiliation{%
 Department of Physics, Dibrugarh University, Dibrugarh \\
 Assam, India, 786004}%
 \affiliation{Theoretical Physics Division, Centre for Atmospheric Studies, Dibrugarh University, Dibrugarh, Assam, India 786004}


\begin{abstract}
We study the nature of particle geodesics around a non-linear electrodynamic black hole (NLED-BH) inspired by the confinement of a heavy quark-antiquark system, which reduces to Maxwell's linear electrodynamics theory at the strong field regime. The corrected BH solution is a special generalisation of the Schwarzschild BH at the linear regime. Such a type of corrected system is parameterised by a charge parameter along with a non-linear electrodynamic term $\zeta$. To be specific, we studied the geodesic behaviour of massless null particles through the geodesic equations using the backward ray-tracing method. We also investigated how NLED effects in charged BH spacetimes affect timelike particle orbits, specifically properties like precession frequency and orbital velocity around the NLED BH. Furthermore, to extend the analysis to the rotating case, we used the modified Newman-Janis algorithm to generate the rotating analogue of the NLED BH. We then analysed the ergosphere formation and shadow cast by the rotating analogue of the NLED BH.
\end{abstract}

\maketitle
%\newpage
%\tableofcontents

\section{Introduction}
Black holes (BHs) are fascinating and enigmatic entities in the universe from both a theoretical and observational perspective. Understanding how massive and massless particles behave in the BH environment is very important in studying the gravitational effects of black holes. Particle geodesics, or the paths that particles follow through spacetime, are an invaluable tool for studying the gravitational environment surrounding black holes.  Additionally, an understanding of the processes, like accretion disk dynamics, gravitational lensing, and black hole shadow formation—all of which produce observable signals that could be used to identify and characterize BHs—requires an understanding of geodesics.
 
Maxwell's equations govern the linear theory of classical electrodynamics. This linearity might not be true near very strong electromagnetic fields, particularly those found close to BHs. It may be noted that strong-field effects are incorporated into Maxwell's theory through the use of nonlinear electrodynamics. The NLED effects in the BH spacetime become important for several reasons. These include the possibility of observational signatures in phenomena like gravitational lensing and BH shadows, NLED as an emergent phenomenon of an effective theory from quantum electrodynamics or string theory, and the possible resolution of central singularities through the resulting modified spacetime geometry. 
 
NLED-BHs exhibit several key features and properties that distinguish them from classical black holes. The spacetime geometry around NLED BHs is modified due to the non-linear interaction between gravity and electromagnetism. This can affect the location and number of event horizons, the structure of the ergosphere (for rotating NLED BHs), and the behaviour of particle geodesics. The energy conditions, which constrain the types of matter allowed in general relativity, may be violated in some NLED models. NLED black holes can possess magnetic charges, leading to unique properties related to their interaction with magnetic fields. Certain NLED models lead to regular black hole solutions, devoid of singularities \cite{Burinskii2002May,Guerrero2020Jul,DeFelice2025Mar,
DeFelice2025Feb,Bronnikov2023Jul,
Bronnikov2001Jan,Bronnikov2018Jan}. The thermodynamics of NLED black holes can also be significantly different from those of classical black holes, with modified temperature, entropy, and heat capacity relations\cite{Sucu2025Jun,Ali2025Jan,
Kruglov2025Apr,Sekhmani2025Mar,Rayimbaev2025Mar,
Gursel2025Mar,Hamil2025Mar}. 
The study of NLED black holes has attracted significant attention in recent years. Born-Infeld electrodynamics is one of the earliest and most well-known NLED theories \cite{Max1934Mar,Born1933Dec}, where NLED effects can arise as limiting cases of certain string theories \cite{Tseytlin2000Jul}. Euler-Heisenberg electrodynamics arises as an effective theory from QED, accounting for quantum corrections to Maxwell's equations\cite{Guerrero2020Jul,
Stefanov2007Jun,Breton2022May,
Amaro2020Nov}. Other NLED models include the exponential NLED \cite{Panotopoulos2019Jun,
Sheykhi2014Aug}, logarithmic NLED \cite{Gullu2021Jun,Mazharimousavi2009Jul}, and power-law NLED \cite{Sorokin2022Aug}. The impact of NLED on gravitational lensing and black hole shadows has been extensively studied in \cite{Junior2025Mar,
Sharipov2025Apr,Waseem2025Feb,
Al-Badawi2025Apr}. For example, it has been shown that the presence of non-linear electromagnetic fields alters the photon sphere radius as well as the critical impact parameter corresponding to the photon orbit, leading to measurable differences in light bending. Furthermore, the effects of NLED on the orbital velocity of stars that orbit a magnetic black hole have been investigated, with the observation that the NLED parameter can reduce the effects of the electric charge at large distances \cite{Mazharimousavi2024Feb}. These studies shed light on the potential of using astrophysical observations to probe NLED effects as well as testing the validity of different NLED models \cite{Allahyari2020Feb}.

Unlike their non-rotating counterparts, rotating black holes have some additional special characteristics. One of such features is the existence of the ergosphere, which is a region outside the event horizon and whose size varies with rotational intensity. Objects are compelled to co-rotate with the black hole inside the ergosphere. Another such feature of rotating black holes is the dragging of spacetime, sometimes referred to as the Lense-Thirring effect, in which the motion of objects nearby is influenced by the black hole's rotation. Rotating black holes are also the only ones with an inner horizon in addition to the event horizon.
The ergosphere is the region where space-time is dragged so strongly that an object can't remain stationary from the point of view of an observer located at infinity. This dragging effect is a direct consequence of the black hole's rotation. The Penrose process, a theoretical mechanism for extracting energy from a rotating black hole, relies on the existence of the ergosphere. Frame-dragging affects the motion of particles and light around the black hole, causing them to precess and deviate from their expected paths. This effect has been experimentally verified in the weak-field limit by observing the precession of gyroscopes in Earth's orbit.

Recently, there have been several interesting works on how the NLED effects play a role in rotating BHs. For instance,
Khoshrangaf et al \cite{Khoshrangbaf2025Jul} investigated how the properties of accretion disks vary when the surrounding regular black holes are influenced by nonlinear electrodynamics. By comparing these black holes with a few conventional BHs, to estimate the observable features such as the disk's inner stable orbit and energy conversion efficiency, and linking these effects to the black hole's charge and spin parameters. The optical properties of a novel BH solution resulting from GR coupled with nonlinear electrodynamics were investigated by Raza et al. \cite{Raza2024May}. After calculating the angular velocity and instability of photon orbits in a spherically symmetric spacetime, they employed the Newman–Janis algorithm to explore the rotating counterpart. By utilizing the Event Horizon Telescope (EHT) observational data, they constrained the model parameters and examined the characteristics of the photon sphere and hence the shadow, noting how spin, charge, and nonlinearity change its shape.
Another important work on the usage of the Newman-Janis algorithm to NLED BHs was carried out by Ghosh and Walia \cite{Ghosh2021Nov}, where they used a modified Newman-Janis algorithm to extend their exact spherically symmetric magnetically charged black hole solution in GR coupled to nonlinear electrodynamics to a rotating BH.

Motivated by these aforementioned works, we undertake the investigation of the null geodesics around the NLED BH in GR. Moreover, we extend the work to the rotating case where we apply the Newman-Janis algorithm to generate the rotating analogue of the NLED BH. In section \ref{sec2}, we revisit the theoretical formulation of the NLED BH to be studied in this text. In section \ref{sec3}, we develop the geodesic equations to obtain the null and timelike geodesic behaviour around the NLED BH. In section \ref{sec4}, we apply the Newman-Janis algorithm to the non-rotating NLED BH to obtain the rotating version of it and study the ergoregion and its spin dependence. Finally, in section \ref{sec5} we solved the Hamilton-Jacobi equations to obtain the shadow of the rotating NLED and its relation to the model parameters and finally in Section \ref{conc_sec}, we summarize the study with the conclusion.

\section{Brief Review of the NLED Black Hole}
\label{sec2}
In this section, we present an overview of black hole solutions arising from Einstein’s non-linear electrodynamics (NLED), which was introduced by Mazharimousavi recently \cite{Mazharimousavi2024Feb}. To construct the model in Eq.~(17) of Ref. \cite{Mazharimousavi2024Feb}, the authors were motivated by the need to address a gap in existing NLED frameworks. Conventional NLED models such as those of Born--Infeld and Euler--Heisenberg recover Maxwell's theory in the weak-field regime but exhibit nonlinear corrections at strong fields. In contrast, for modelling confinement--- as seen in the interaction between heavy quark--antiquark pairs. A theory is desired that behaves linearly in the strong-field (short-distance) regime while inducing confining dynamics at weak fields (large distances). Mazharimousavi reverse-engineered a modified Lagrangian of the form $\mathcal{L}_{e} = -\varepsilon(F)F$, where $\varepsilon(F)$ is given by Eq. \eqref{eq:4}, which was inspired by Guendelman's square-root model \cite{Guendelman2004Jul}, that generates a Cornell-type potential \cite{Eichten1975Feb} but lacks the proper ultraviolet behavior. They derive the nonlinear function $\varepsilon(F)$ given in Eq. (17) of \cite{Guendelman2004Jul} by prescribing an electric field profile that combines a Coulomb term with a confining component, $E(r) = q/r^{2} + \zeta \sqrt{q/r}$, and solving the generalized Maxwell equations. By combining Maxwellian and confining dynamics into a single framework, this formulation guarantees the proper limiting behavior in both the strong- and weak-field regimes. Subsequently, the model was investigated very recently by Waseem et al. \cite{Waseem2025Feb} to study gravitational lensing properties.

 The action in this model is expressed as
\begin{equation}
    S = \int d^{4}x \sqrt{-g} \left( \frac{\mathcal{R}}{16\pi} + \mathcal{L}\left(\mathcal{F}\right) \right), \label{eq:2}
\end{equation}
where $\mathcal{R}$ denotes the Ricci scalar, the gravitational constant is set to $G=1$, and the Lagrangian density $\mathcal{L}\left(\mathcal{F}\right)$ is defined as  \cite{Mazharimousavi2024Feb,Waseem2025Feb}
\begin{equation}
    \mathcal{L}=-\varepsilon\left(\mathcal{F}\right)\mathcal{F}, \label{eq:3}
\end{equation}
with
\begin{equation}
    \varepsilon\left(\mathcal{F}\right) = \frac{16\left(3\sqrt{2\mathcal{F}} + \zeta \left(\zeta + \sqrt{\zeta^2 + 4\sqrt{2\mathcal{F}}}\right)\right)}{3\left(\zeta + \sqrt{\zeta^2 + 4\sqrt{2\mathcal{F}}}\right)^4}. \label{eq:4}
\end{equation}
Here, $\zeta$ is a positive constant parameter. In the limit of a small $\zeta$, one finds that
\begin{equation}
    \varepsilon\left(\mathcal{F}\right)  \simeq 1-\frac{4}{3}\frac{\sqrt[4]{2}}{\sqrt[4]{\mathcal{F}}}\zeta+\left(\frac{\sqrt[4]{2}}{\sqrt[4]{\mathcal{F}}}\right)^{2}\zeta^2+\mathcal{O}\left(\zeta^{3}\right). \label{eq:5}
\end{equation}
In this context, $\mathcal{F}$ stands for the electromagnetic invariant, $\frac{1}{4}F_{\mu \nu}F^{\mu \nu}$. Now, varying the action in Eq. (\ref{eq:2}) with respect to the NLED energy-momentum tensor yields the Einstein-NLED field equations
\begin{equation}
    G^{\nu}_{\mu}=8\pi T^{\nu}_{\mu},\label{eq:6}
\end{equation}
where the energy-momentum tensor is given by
\begin{equation}
    T_{\mu}^{\nu}=\frac{1}{4\pi}\left(\mathcal{L}\delta_{\mu}^{\nu}-\mathcal{L}_{\mathcal{F}}\mathcal{F}_{\mu \lambda}\mathcal{F}^{\nu \lambda}\right). \label{eq:7}
\end{equation}
Furthermore, if one varies the action with reference to the gauge fields results in the NLED version of Maxwell's equations
\begin{equation}
    d\left(\mathcal{L}_{\mathcal{F}} \tilde{\mathbf{F}}\right) = 0, \label{eq:8}
\end{equation}
where the dual electromagnetic field tensor is defined via
\begin{equation}
    \textbf{F}=\frac{1}{2}F_{\mu \nu }dx^{\mu} \wedge dx^{\nu}. \label{eq:9}
\end{equation}
Assuming the presence of a magnetic monopole at the origin, the corresponding field configuration is adopted as
\begin{equation}
    \textbf{F}=Q \,\sin\theta \, d\theta \wedge d\phi, \label{eq:10}
\end{equation}
With $Q>0$ denoting the magnetic charge. If we choose to work with the spherically symmetric solution for the NLED black hole given by
\begin{equation}
	ds^2 = -f(r)dt^2 + \frac{1}{f(r)} dr^2 + r^2 d\Omega^2,
	\label{metric_ned}
\end{equation} 
The $tt$-component of Eq. (\ref{eq:7}) then reduces to
\begin{equation}
    \frac{rf^{\prime}\left(r\right)+f\left(r\right)-1}{r^{2}}=-\frac{4B^{3}\left(2\zeta^{2}+2\zeta\sqrt{\zeta^{2}+2B}+3B\right)}{3\left(\zeta+\sqrt{\zeta^{2}+2B}\right)^{4}}, \label{eq:11}
\end{equation}
where $B=\frac{Q}{r^{2}}$ represents the radial dependence of the magnetic field due to the monopole.
The complete solution of Eq. (\ref{eq:11}) has been obtained in \cite{Mazharimousavi2024Feb} and is given by
\begin{equation}
\begin{aligned}
    f\left(r\right)&=1-\frac{2M}{r}+\frac{Q^{2}}{r^{2}}-2Q\zeta^{2}-\frac{2\zeta^{4}r^{2}}{9}\\&-\frac{4}{3}\frac{Q^{\frac{3}{2}}\zeta \sqrt{2}}{r}\times \text{ln}\left(\frac{4Q+2\sqrt{2Q}\sqrt{\zeta^{2}r^{2}+2Q}}{r}\right)\\&+\frac{2\zeta \sqrt{\zeta^{2}r^{2}+2Q}\left(\zeta^{2}r^{2}+8Q\right)}{9r}.
\end{aligned}
\label{eq:12}
\end{equation}
This metric function satisfies all of the Einstein field equations. In the limit $\zeta \to 0$, it naturally reduces to the standard Reissner-Nordstr\"om solution,
\begin{equation}
    f\left(r\right)=1-\frac{2M}{r}+\frac{Q^{2}}{r^{2}}, \label{eq:13}
\end{equation}
whereas for small $\zeta$, the behavior of $f\left(r\right)$ is characterized by
\begin{equation}
\begin{aligned}
    f\left(r \right)&=1-\frac{2M}{r}+\frac{Q^{2}}{r^{2}}+\zeta \left(\frac{4Q^{3/2}\sqrt{2}\left(4-3\text{ln}\left(\frac{8Q}{r}\right)\right)}{9r}\right)\\& \hspace{4cm}+\mathcal{O}\left(\zeta^{2}\right).
    \end{aligned} \label{eq:14}
\end{equation}
The asymptotic form of the metric function exhibits notable features, being expressible as
\begin{equation}
    f\left(r\right)=1-\frac{2M_{\rm ADM}}{r}+\frac{Q^{3}}{9\zeta^{2}r^{4}}+\mathcal{O}\left(\frac{1}{r^{6}}\right), \label{eq:15}
\end{equation}
with the Arnowitt–Deser–Misner (ADM) mass defined by
\begin{equation}
    M_{\rm ADM} = M + \frac{2\sqrt{2}}{3} \zeta Q^{3/2} \ln\left(2\zeta \sqrt{2Q}\right). \label{eq:16}
\end{equation}
 One can notice from Eq. \eqref{eq:14} that the term $Q^3/9\zeta^2 r^4$ decays away very rapidly with radial distances compared to the second term of the equation. Therefore, we can safely ignore this term. This further simplifies the model. Thus the lapse function $f(r)$ takes the form:
\begin{equation}
f(r) =1 -\frac{2 M}{r}-\frac{4 \sqrt{2} \zeta  Q^{3/2} \log \left(2 \sqrt{2} \zeta  \sqrt{Q}\right)}{3 r}.
\label{lapse1}
\end{equation}
It can be easily verified that as $\zeta \to 0$ and $Q \to 0$, the metric resembles the standard Schwarzschild metric. 

\begin{figure*}[htb]
\centerline{\includegraphics[scale=0.6]{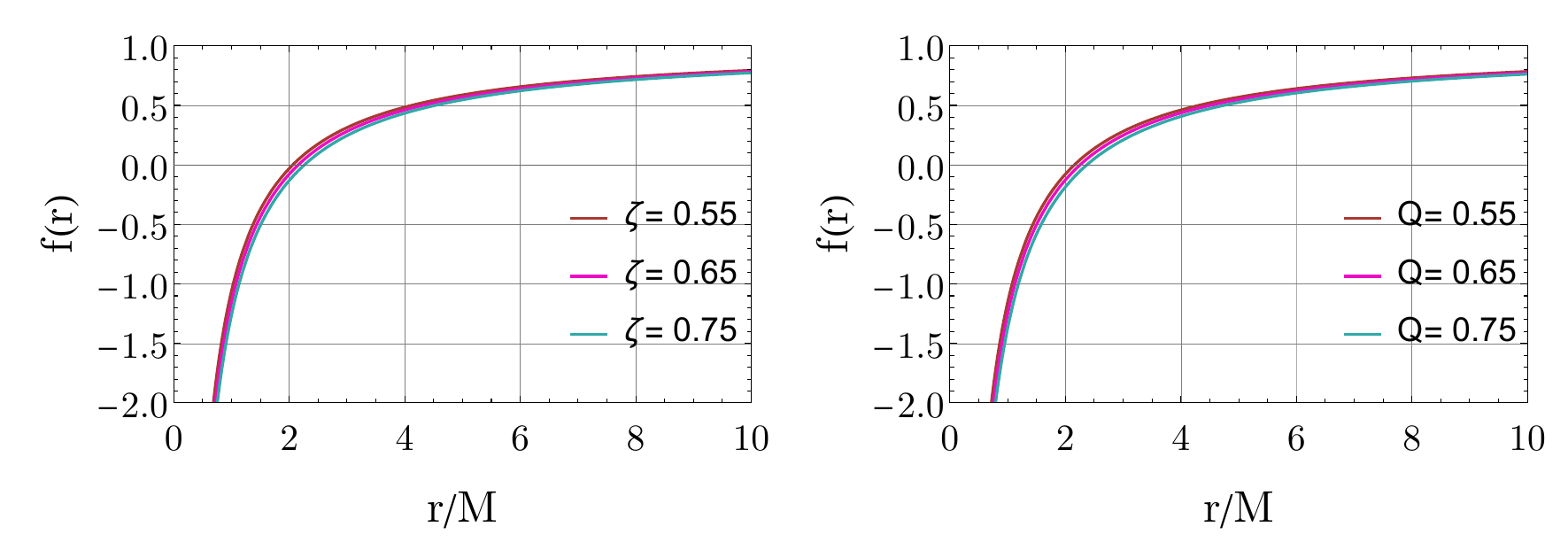}}
\centerline{\hspace{5cm} (a) \hspace{8.5cm} (b) \hspace{3cm}}
\caption{The lapse function is shown with respect to different values of $\zeta$ and $Q$}
\label{lapse_plot}
\end{figure*}

Fig. (\ref{lapse_plot}) depicts the behavior of the lapse function. The event horizons are identified where \( f(r) = 0 \). It clearly shows that there is a single horizon for a given value of $\zeta$ and $Q$. As non-linearity $\zeta$ and charge $Q$ increase, the horizon is shifted towards a larger value.

\section{Geodesic Behaviour}
\subsection{Null Geodesics}
\label{sec3}

In spherically symmetric spacetimes, the metric functions \(A(r)\) and \(B(r)\) serve as the redshift factor and the measure of spatial curvature, respectively, thereby governing the dynamics of both massless and massive particles within the gravitational region of a black hole. The metric admits two intrinsic Killing vectors, corresponding to time translation and rotational symmetry, which lead to conserved quantities along geodesics. Specifically, the Killing vector associated with time translation yields a conserved energy, 
\[
E = A(r)\dot{t},
\]
which remains invariant along the particle’s geodesic. Similarly, the Killing vector corresponding to rotational symmetry gives rise to a conserved angular momentum that constrains the motion to orbital trajectories---whether parabolic, hyperbolic, or elliptical, depending on the eccentricity. These symmetries and the associated conserved quantities play a crucial role in formulating the equations of motion. Furthermore, one may derive the effective potential governing the trajectories of both null and timelike particles to determine orbital stability and the possible occurrence of orbital precession for timelike geodesics.

To derive the geodesic equations of motion, we begin with the general form of a spherically symmetric spacetime metric:
\begin{equation}
ds^2 = - A(r)^2 dt^2 + B(r)^2 dr^2 + r^2 d\theta^2 + r^2 \sin^2 \theta d\phi^2,
\label{gen_met} 
\end{equation}
which possesses time translational and rotational symmetries. Consequently, the Killing vectors associated with these symmetries yield conserved quantities along the geodesics, expressed as
\begin{equation}
K_\mu \dot{x}^\mu = \text{ constant}.
\label{kill}
\end{equation}
Here, the notation \(\dot{x}^\mu\) indicates differentiation with respect to the affine parameter \(\lambda\). The Killing vectors corresponding to time translational and rotational symmetry are given by
\begin{equation}
K_\mu = (-A(r),\, 0,\, 0,\, 0),
\label{time_killing}
\end{equation}
and 
\begin{equation}
K_\mu = (0,\, 0,\, 0,\, r^2 \sin^2 \theta),
\label{sph_killing}
\end{equation}
respectively.

From Eqs. \eqref{gen_met} and \eqref{time_killing}, one obtains the energy equation:
\begin{equation}
E = A(r)\dot{t} = \text{constant}.
\label{E-eqn}
\end{equation}
Exploiting the rotational symmetry allows the analysis to be confined to the equatorial plane by setting \(\theta = \pi/2\). This restriction leads to the angular momentum equation from Eqs. \eqref{gen_met} and \eqref{sph_killing}:
\begin{equation}
L = r^2 \dot{\phi} = \text{constant},
\label{L-eqn}
\end{equation}
where \(L\) denotes the conserved angular momentum.

Moreover, the norm of the tangent vector to the geodesic is conserved, which is expressed by
\begin{equation}
\epsilon = - g_{\mu \nu} \dot{x}^\mu \dot{x}^\nu,
\label{eps}
\end{equation}
with \(\epsilon = 0\) for null geodesics and \(\epsilon = 1\) for timelike geodesics. Substituting the metric \eqref{gen_met} into Eq. \eqref{eps} gives
\begin{equation}
-\epsilon = -A(r)\dot{t}^2 + B(r)\dot{r}^2 + r^2 \dot{\phi}^2.
\label{eps1}
\end{equation}
Equation \eqref{eps1} may alternatively be rearranged as
\begin{equation}
\dot{r}^2 = \frac{E^2}{A(r)B(r)} - \frac{L^2}{r^2B(r)} - \frac{\epsilon}{B(r)}.
\label{rdotsq}
\end{equation}

To derive the geodesic equations, one may employ the Lagrangian defined by
\begin{equation}
\mathcal{L} = \frac{1}{2} g_{\mu \nu} \dot{x}^\mu \dot{x}^\nu = \frac{1}{2}\left(-A(r)\dot{t}^2 + B(r)\dot{r}^2 + r^2\dot{\phi}^2\right).
\label{Lag}
\end{equation}
Application of the Euler–Lagrange equation in the \(r\)-coordinate,
\begin{equation}
\frac{d}{d\lambda}\left( \frac{\partial \mathcal{L}}{\partial \dot{r}}\right) = \frac{\partial \mathcal{L}}{\partial r},
\label{Lagr}
\end{equation}
yields
\begin{equation}
\dot{p}_r = \frac{1}{2}\left(- \frac{\partial A(r)}{\partial r}\dot{t}^2 + \frac{\partial B(r)}{\partial r}\dot{r}^2 + 2r\dot{\phi}^2 \right),
\label{prdot}
\end{equation}
where the conjugate momentum associated with the \(r\)-coordinate is defined as
\begin{equation}
p_r = \frac{\partial \mathcal{L}}{\partial \dot{r}} = \dot{r}B(r).
\label{pr}
\end{equation}

By combining Eqs. \eqref{E-eqn}, \eqref{L-eqn}, \eqref{prdot}, and \eqref{pr}, the equations of motion for null geodesics in the general spherically symmetric spacetime \eqref{gen_met} can be expressed as:
\begin{equation}
\begin{aligned}
\dot{t} &= E\, A(r)^{-1}, \\
\dot{\phi} &= \frac{L}{r^2}, \\
\dot{r} &= p_r\, B(r)^{-1}, \\
\dot{p}_r &= \frac{1}{2}\left(- \frac{E^2}{A(r)^2}\frac{\partial A(r)}{\partial r} + \frac{p_r^2}{B(r)^2}\frac{\partial B(r)}{\partial r} + \frac{2L^2}{r^3}\right).
\end{aligned}
\label{geod_eqs}
\end{equation}
Furthermore, from Eq. \eqref{rdotsq}, one may deduce
\begin{equation}
\frac{1}{2}\dot{r}^2 + V_{eff} = \frac{E^2}{2},
\label{eff1}
\end{equation}
with the effective potential defined as
\begin{equation}
V_{eff} =  \frac{L^2}{2r^2}B(r)^{-1} + \frac{\epsilon}{2B(r)}.
\label{eff2}
\end{equation}

By employing the metric given by \(\eqref{lapse1}\) and comparing it with the metric \eqref{gen_met}, one finds that \(A(r) = f(r)\) and \(B(r) = f(r)^{-1}\). Consequently, the system of Eqs. \eqref{geod_eqs} assumes the form
\begin{equation}
\begin{aligned}
\dot{t} &= E \left(1 -\frac{2M}{r}-\frac{4\sqrt{2}\,\zeta\,Q^{3/2}\log\left(2\sqrt{2}\,\zeta\,\sqrt{Q}\right)}{3r}\right)^{-1}, \\
\dot{\phi} &= \frac{L}{r^2}, \\
\dot{r} &= p_r \left(1 -\frac{2M}{r}-\frac{4\sqrt{2}\,\zeta\,Q^{3/2}\log\left(2\sqrt{2}\,\zeta\,\sqrt{Q}\right)}{3r}\right), \\
\dot{p}_r &= -\frac{3E^2\left(3M+2\sqrt{2}\,\zeta\,Q^{3/2}\log\left(2\sqrt{2}\,\zeta\,\sqrt{Q}\right)\right)}{\left(6M+4\sqrt{2}\,\zeta\,Q^{3/2}\log\left(2\sqrt{2}\,\zeta\,\sqrt{Q}\right)-3r\right)^2} \\
&\quad -\frac{p_r^2\left(3M+2\sqrt{2}\,\zeta\,Q^{3/2}\log\left(2\sqrt{2}\,\zeta\,\sqrt{Q}\right)\right)}{3r^2}+\frac{L^2}{r^3}.
\end{aligned}
\label{geod_eqs_mod}
\end{equation}

\begin{figure}
	\centerline{\includegraphics[scale=0.35]{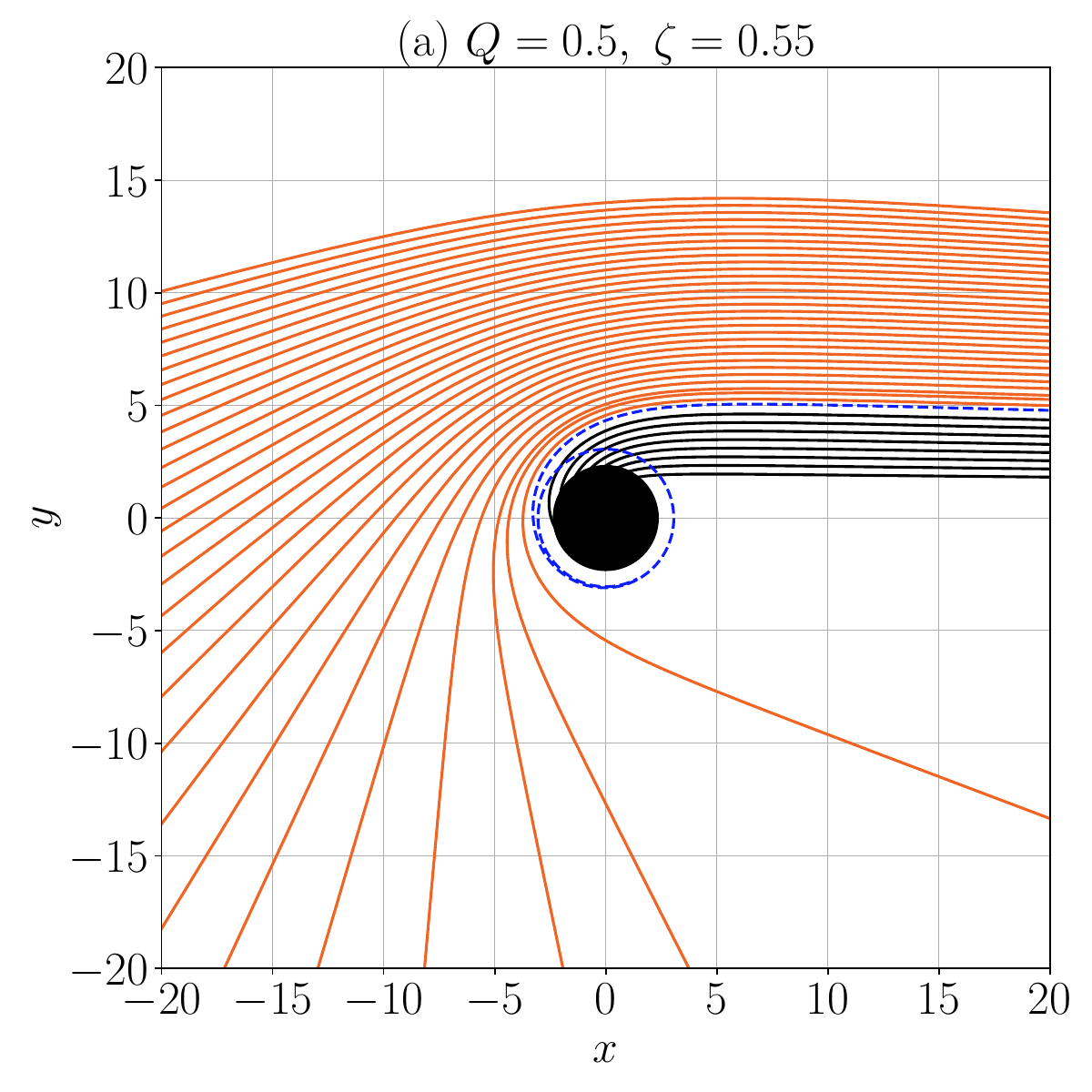}}
	\centerline{\includegraphics[scale=0.35]{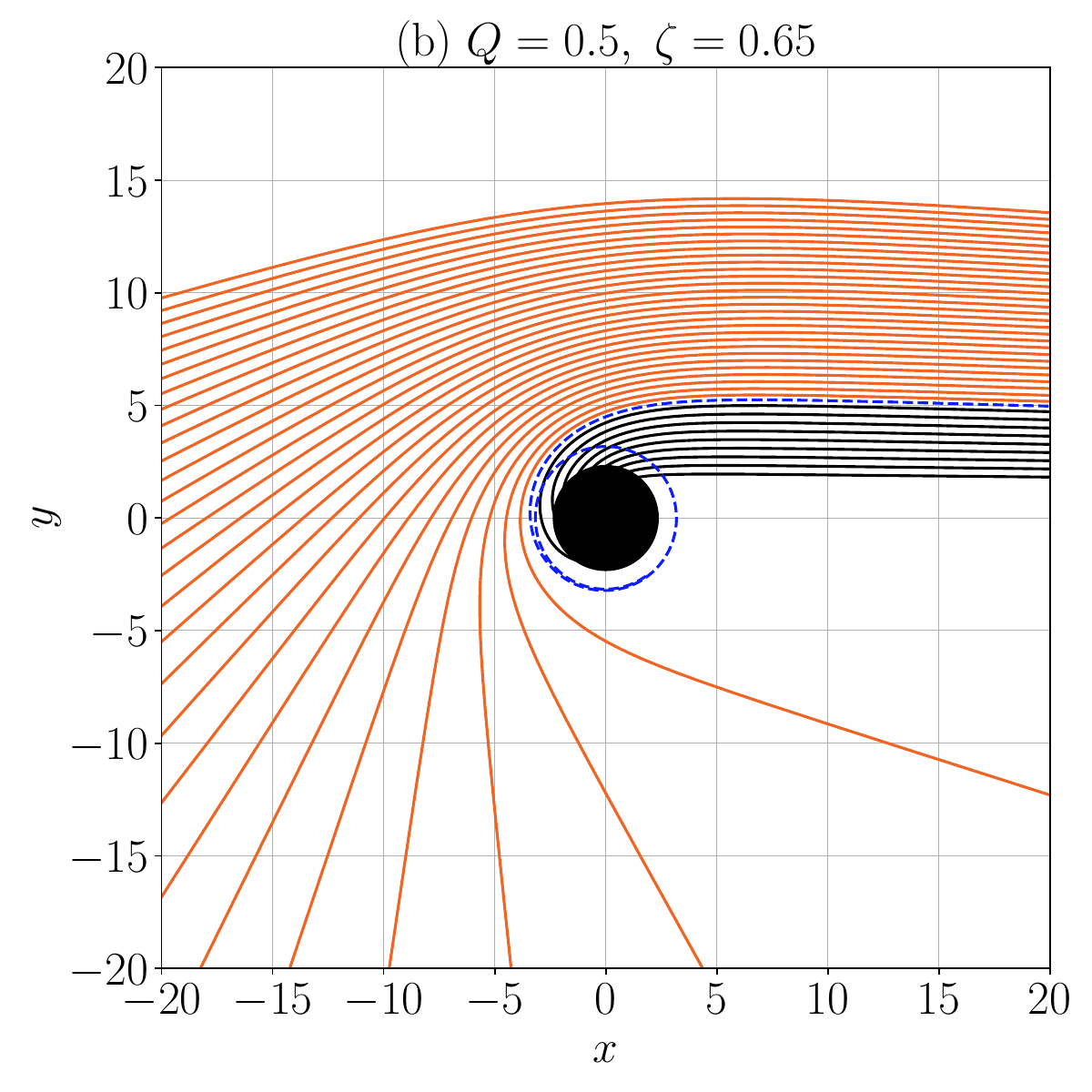}}
	\centerline{\includegraphics[scale=0.35]{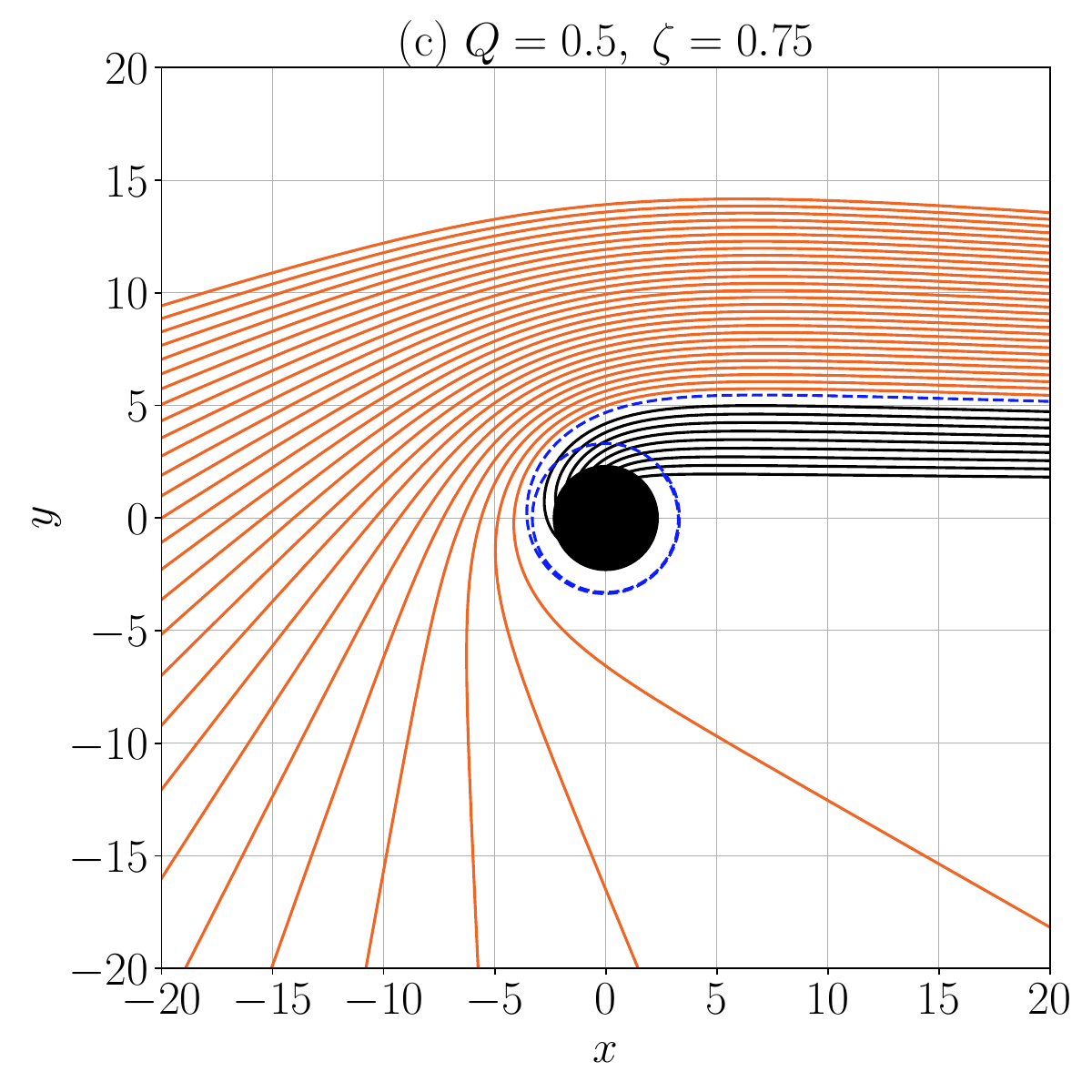}}
	\caption{The null trajectories obtained through the backward ray-tracing around the NLED BH are shown for different values of $\zeta$.}
	\label{null_brt_zeta}
\end{figure}

\begin{figure}
	\centerline{\includegraphics[scale=0.35]{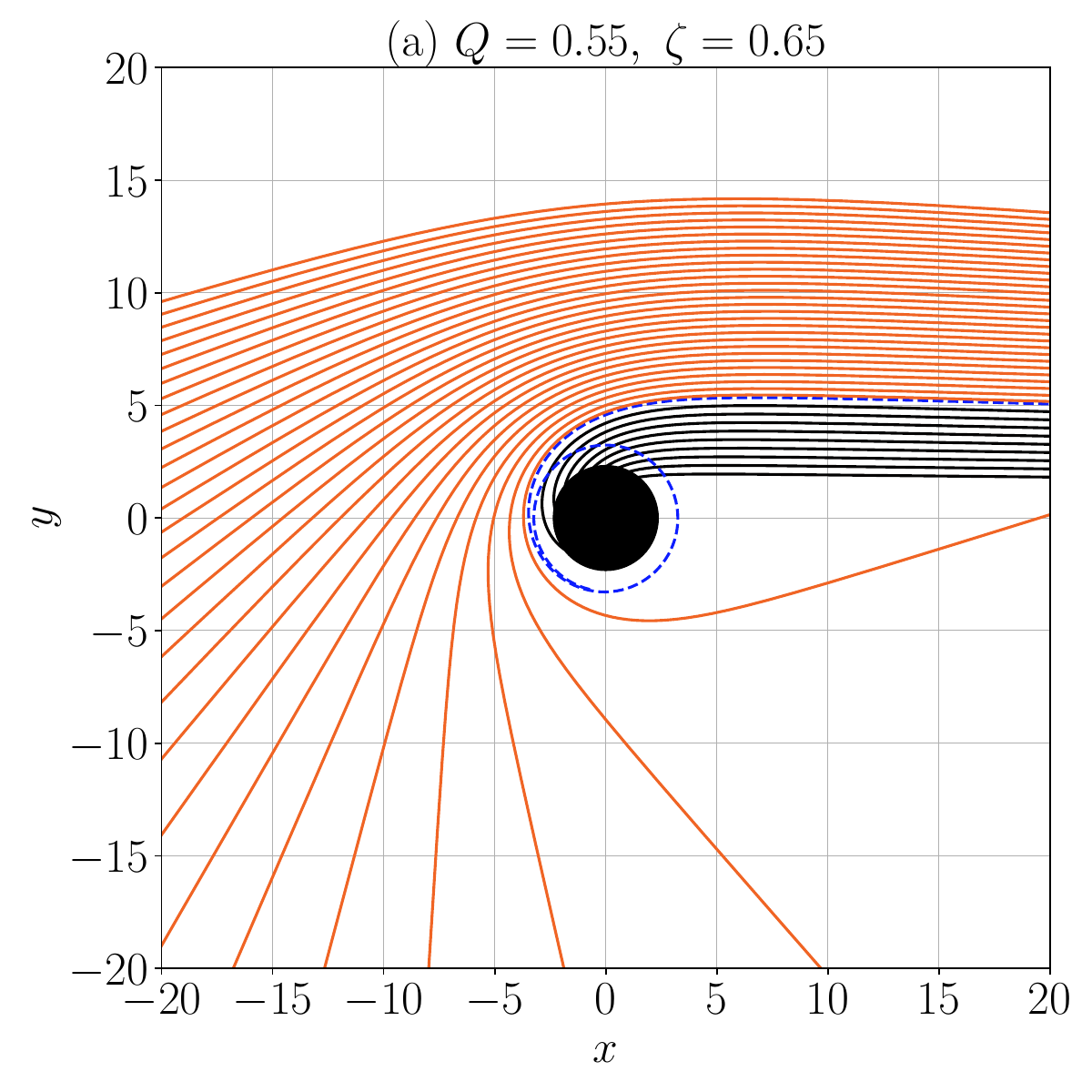}}
	\centerline{\includegraphics[scale=0.35]{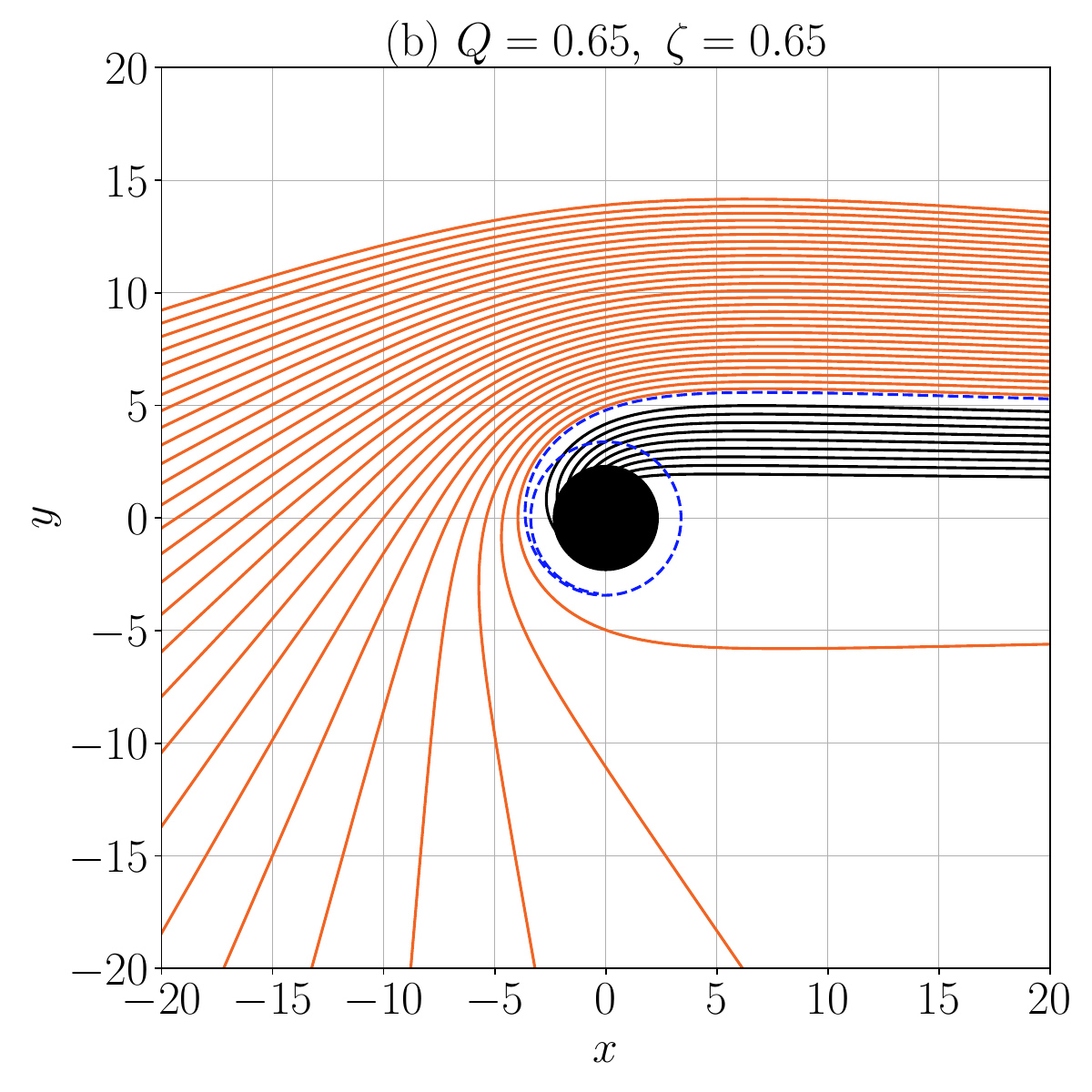}}
	\centerline{\includegraphics[scale=0.35]{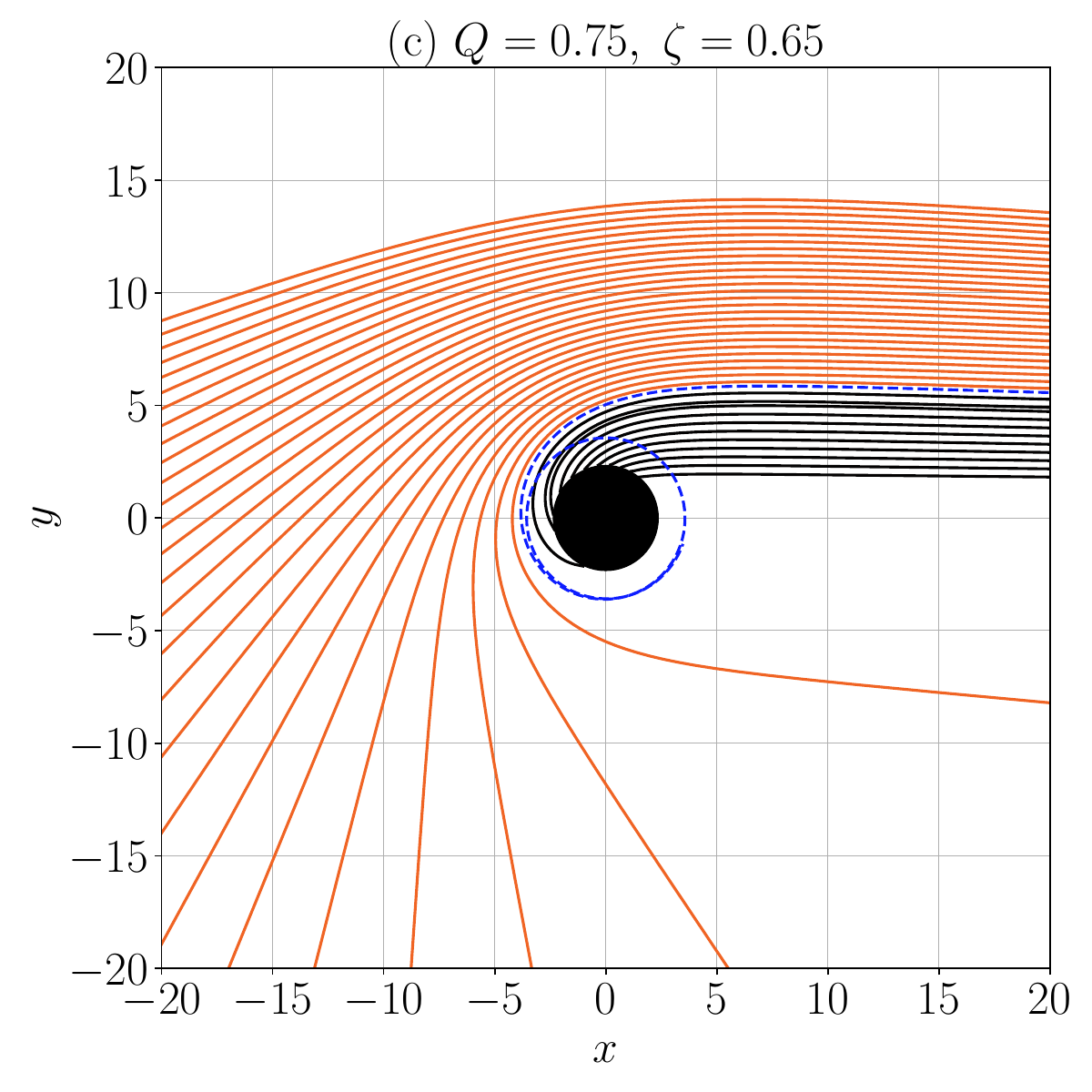}}
	\caption{The null trajectories obtained through the backward ray-tracing around the NLED BH are shown for different values of $Q$.}
	\label{null_brt_Q}
\end{figure}

Furthermore, the effective potential for null geodesics in the present model is computed from Eq. \eqref{eff2} as
\begin{equation}
V_{eff} = -\frac{L^2M}{r^3}-\frac{2\sqrt{2}\,\zeta\,L^2Q^{3/2}\log\left(2\sqrt{2}\,\zeta\,\sqrt{Q}\right)}{3r^3}+\frac{L^2}{2r^2},
\label{eff_pot}
\end{equation}
with \(\epsilon\) fixed to \(0\) for null geodesics. The null geodesic trajectories are obtained by numerically solving the set of equations given in Eq. \eqref{geod_eqs_mod}, as depicted in Figs. (\ref{null_brt_zeta}) and (\ref{null_brt_Q}). In this figure, the circular photon orbits are highlighted by blue dashed curves, with their radii corresponding to the peaks of the effective potential.

\begin{table*}[htb]
	\begin{tabular}{|ccc|ccc|}
		\hline
		\multicolumn{3}{|c|}{$Q = 0.5$}                                          & \multicolumn{3}{c|}{$\zeta = 0.65$}                                         \\ \hline
		\multicolumn{1}{|c|}{$\zeta$} & \multicolumn{1}{c|}{$r_{ph}$} & $L$      & \multicolumn{1}{c|}{$Q$}  & \multicolumn{1}{c|}{$r_{ph}$} & $L$      \\ \hline
		\multicolumn{1}{|c|}{0.55}    & \multicolumn{1}{c|}{3.050451} & 5.255096 & \multicolumn{1}{c|}{0.55} & \multicolumn{1}{c|}{3.248265} & 5.575755 \\ \hline
		\multicolumn{1}{|c|}{0.65}    & \multicolumn{1}{c|}{3.157110} & 5.469900 & \multicolumn{1}{c|}{0.65} & \multicolumn{1}{c|}{3.373954} & 5.826450 \\ \hline
		\multicolumn{1}{|c|}{0.75}    & \multicolumn{1}{c|}{3.250431} & 5.700580 & \multicolumn{1}{c|}{0.75} & \multicolumn{1}{c|}{3.639472} & 6.127820 \\ \hline
	\end{tabular}
	\caption{The table shows the position of the photon sphere with their corresponding angular momentum values for different choices of $\zeta$ and $Q$. Here energy of the photon $E$ is set to 1.}
	\label{tab_null}
\end{table*}

We analyze null geodesics around a spherically symmetric black hole modified by nonlinear electrodynamics and find that the photon-sphere radius is systematically shifted outward by the coupling parameter~$\zeta$ and the electric charge~$Q$ as displayed in Tab. \ref{tab_null}. The critical photon orbit increases from $r_{ph}\approx 3.05$ to $3.25$ at fixed $Q=0.5$ when $\zeta$ is increased from 0.55 to 0.75, and the corresponding angular momentum increases simultaneously. Likewise, for $\zeta=0.65$, increasing the $Q$ values from $0.55$ to $0.75$ causes the photon orbit radius $r_{ph}$ to increase from $3.25$ to $3.64$. Possibly, the Coulombic term modifies the effective potential more than the NLED correction does, and the sensitivity to charge is stronger. These findings suggest that a larger photon sphere is produced by both increased charge and improved nonlinear electrodynamics.

\subsection{Timelike Geodesics}
According to general relativity, the spacetime curvature causes the precession of perihelion around a black hole for massive particles. This is caused by relativistic processes, particularly those that occur close to the black hole.  With each loop, the massive particles (timelike geodesics) orbiting the black hole (BH) in its gravitational field advance the perihelion, the closest point of approach, and undergo a slight angular deviation from a perfect ellipse. 

To study the properties of timelike particles around the NLED BH, we start from the effective potential of the particles, which can be obtained from Eq. \eqref{eff2} as:
\begin{equation}
	\begin{aligned}
	V_{eff} &= \frac{\left(L^2+r^2\right)}{6 r^3}\times\\ &\left[\left(-6 M-4 \sqrt{2} \xi  Q^{3/2} \log \left(2 \sqrt{2} \xi  \sqrt{Q}\right)+3 r\right)\right]
		\end{aligned}
	\label{timelike-eff_pot_eq}
\end{equation}
\begin{figure*}
	\centerline{\includegraphics[scale = 0.5]{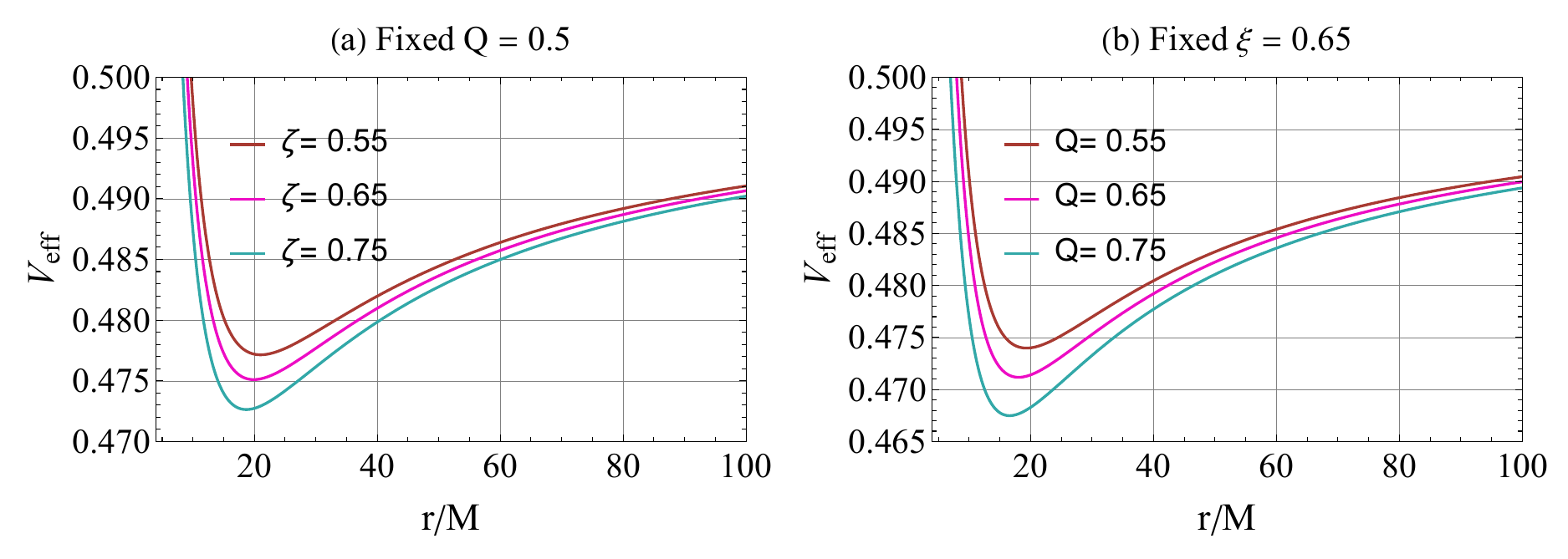}}
	\caption{The effective potential for timelike particles is shown. Here we have set $M = 1$ and $L = 5$.}
	\label{eff_plot_timelike}
\end{figure*}
From the plot of the effective potential, shown in Fig. \ref{eff_plot_timelike} (a) and (b), for timelike orbits, one clearly observes that the well of the effective potential dips towards lower values as the 
NLED parameter $\xi$ and charge $Q$ increase. This implies that as the NLED parameter $\xi$ and electric charge $Q$ increase, the timelike effective potential
\(
V_{\rm eff}(r)
\)
develops a deeper minimum $V_{\min}$ such that the specific energy of a stable bound orbit,
\begin{equation}
E_{\rm bound} = \sqrt{V_{\min}}, \label{cond1_stab}
\end{equation}
decreases (and hence the binding energy increases), while the innermost stable circular orbit (ISCO) radius is determined by
\begin{equation}
\left.\frac{dV_{\rm eff}}{dr}\right|_{r_{\rm ISCO}} = 0
\quad\text{and}\quad
\left.\frac{d^2V_{\rm eff}}{dr^2}\right|_{r_{\rm ISCO}} = 0. \label{cond2_stab}
\end{equation}
 tends to shift inward.  The conditions \eqref{cond1_stab} and \eqref{cond2_stab} therefore correspond to the ISCO for timelike particles.
 
Using Eqs. \eqref{timelike-eff_pot_eq}, \eqref{cond1_stab} and \eqref{cond2_stab} we can derive the angular momentum $L$ and energy $E$ in terms of the radial coordinates as: 
\begin{widetext}
\begin{equation}
\begin{aligned}
L &= \frac{
\sqrt{ -3 M r^2 - 2 \sqrt{2} \zeta Q^{3/2} r^2 \log\left(2 \sqrt{2} \zeta \sqrt{Q} \right) }
}{
\sqrt{3} \sqrt{ 3 M + 2 \sqrt{2} \zeta Q^{3/2} \log\left(2 \sqrt{2} \zeta \sqrt{Q} \right) - r }
}, \\
E &= \pm \frac{1}{3\sqrt{2}} \Biggl[
\frac{
\biggl(
24 \sqrt{2} \zeta Q^{3/2} (2M - r) \log\left(2 \sqrt{2} \zeta Q^{1/2}\right)
+ 9 (r - 2M)^2
+ 32 \zeta^2 Q^3 \log^2\left(2 \sqrt{2} \zeta Q^{1/2} \right)
\biggr)^{1/2}
}{
\biggl[
r \left(
\log\left(2 \sqrt{2} \zeta Q^{1/2}\right)
+ r - 2 \sqrt{2} \zeta Q^{3/2} - 3M
\right)
\biggr]^{1/2}
}
\Biggr]
\end{aligned}
\label{E_L}
\end{equation}

\end{widetext}
The orbital velocity of particles on bound orbits is defined as follows:
\begin{equation}
\begin{aligned}
\Omega &:= \frac{\dot{\phi}}{\dot{r}} = \sqrt{\frac{f'(r)}{2r}}
&= \sqrt{\frac{M+\frac{2}{3} \sqrt{2} \zeta Q^{3/2} \log \left(2 \sqrt{2} \zeta  \sqrt{Q}\right)}{r^3}},
\end{aligned}
\label{orb_vel}
\end{equation}
and the proper angular velocity $\omega = \dot{\phi} = L/r^2$ is related to $\Omega$ through the relation
\begin{equation}
\begin{aligned}
\omega &:= \sqrt{\frac{2}{2f(r) - r f'(r)}} \, \Omega \\
&= \Omega \sqrt{ \frac{r}{ r - 3M - 2\sqrt{2} \zeta Q^{3/2} \log\left(2\sqrt{2} \zeta \sqrt{Q} \right) } }
\end{aligned}
\label{prop_orb_vel}
\end{equation}

Using these definitions, we can next define the geodesic precession frequency $\Theta_{GPF}$ as \cite{Saadati2023Dec}
\begin{equation}
\Theta_{GBF} = \Omega - \Omega_{GBF}, \quad \Omega_{GBF} = \Omega\sqrt{f(r) - \Omega^2 r^2}.
\label{GBF}
\end{equation}
On using Eq. \eqref{orb_vel} in Eq. \eqref{GBF}, we get
\begin{equation}
\begin{aligned}
\Theta_{GBF} &= \sqrt{\frac{f'(r)}{2r}}\left[ 1 - \sqrt{\frac{2f(r) - rf'(r)}{2}}\right] \\
&= \frac{1}{\sqrt{3}} \sqrt{ \frac{3M + 2\sqrt{2} \zeta Q^{3/2} \log\left(2\sqrt{2} \zeta \sqrt{Q} \right)}{r^3} } \times \\ &
\left(1 - \sqrt{ \frac{r -3M - 2\sqrt{2} \zeta Q^{3/2} \log\left(2\sqrt{2} \zeta \sqrt{Q} \right)}{r} } \right)
\end{aligned}
\end{equation}

\begin{figure*}[htb]
\centerline{\includegraphics[scale=0.5]{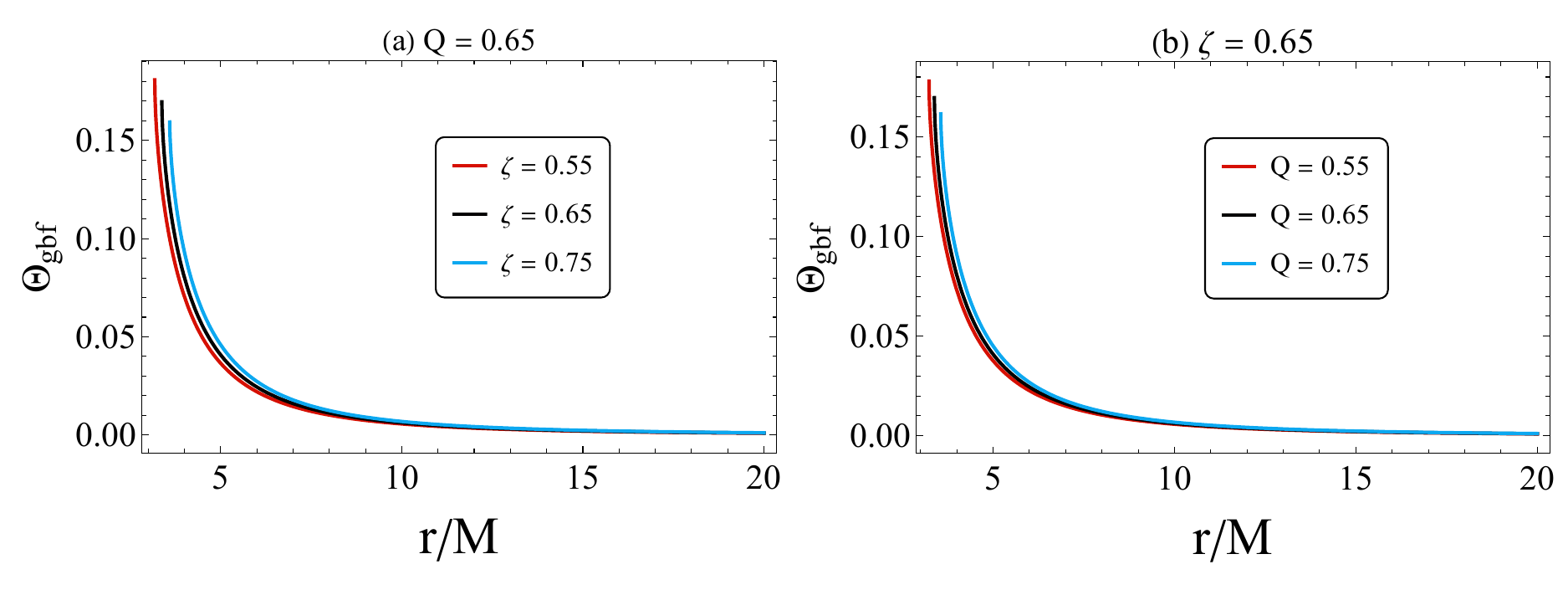}}
\caption{The geodesic precession frequency is shown for different values of the NLED parameter $\zeta$ and charge $Q$}
\label{GPF_plot}
\end{figure*}
From Fig. \ref{GPF_plot}, we see that the geodesic precession frequency \(\Theta_{\rm GPF}(r)\) reaches its maximum at small radial coordinates and monotonically asymptotes towards zero as \(r\) increases in the vicinity of the NLED black hole, indicating the weakening of curvature effects. As we have seen before, stronger NLED corrections deepen the well of the effective potential and thus enhance periastron advance, as evidenced by the relatively small but systematic increase in \(\Theta_{\rm GPF}\) at all radii that occurs when the NLED parameter \(\zeta\) is increased at fixed charge \(Q\). On the other hand, at fixed \(\zeta\), a larger \(Q\) results in a slight decrease in \(\Theta_{\rm GPF}\), which may be attributed to the possible Coulombic effects counteracting the gravitational attraction and thereby reducing orbital precession. Both dependencies become insignificant far from the NLED BH (\(r\gtrsim20M\)), and \(\Theta_{\rm GPF}\) approaches zero, which is pertinent to the Newtonian limit of no per-orbit precession.

Now let us calculate the orbital velocity of the timelike particles around the NLED BH at far away regions from the horizon. One can approximate the lapse function in terms of the gravitational potential for a timelike particle of unit mass as \cite{Saadati2023Dec,Abramowicz2009Jun}
\begin{equation}
f(r) = 1 + 2\Phi(r),
\label{appr_lapse}
\end{equation}
where $\Phi(r)$ denotes the Newtonian gravitational potential. Thus, using our NLED BH metric, we can express $\Phi
(r)$ as 
\begin{equation}
\Phi(r) = -\frac{M+\frac{2}{3} \sqrt{2} \zeta  Q^{3/2} \log \left(2 \sqrt{2} \zeta  \sqrt{Q}\right)}{r}.
\label{Phi_eq}
\end{equation}
This leads to the effective central force through the relation
\begin{equation}
\begin{aligned}
F_c &= - \frac{\partial \Phi(r)}{\partial r} \\&= -\frac{M+\frac{2}{3} \sqrt{2} \zeta Q^{3/2} \log \left(2 \sqrt{2} \zeta  \sqrt{Q}\right)}{r^2}.
\end{aligned}
\label{force}
\end{equation}
For a timelike particle of unit mass, we can relate the centripetal acceleration to the effective force, which gives $\left|F_c\right| = v^2/r$. This implies the orbital velocity can be expressed as
\begin{equation}
v = \sqrt{r} \sqrt{\left| \frac{\frac{2}{3} \xi  \log \left(2 \sqrt{2} \sqrt{Q} \xi \right) \sqrt{2} Q^{3/2}+M}{r^2}\right| }
\label{vel_eq}
\end{equation}

\begin{figure*}[htb]
\centerline{\includegraphics[scale=0.5]{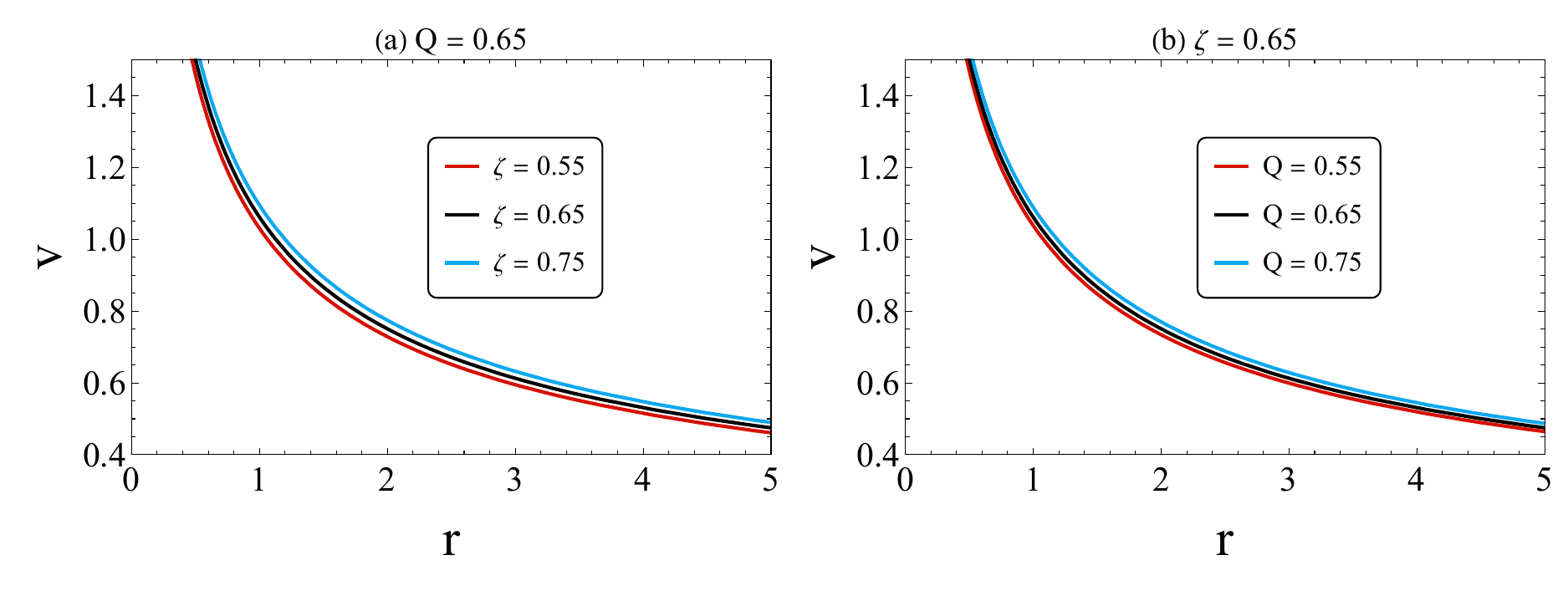}}
\caption{The orbital velocity is plotted against the radial coordinate for different values of NLED parameter $\zeta$ and charge $Q$.}
\label{v_fig}
\end{figure*}
The orbital velocity is plotted for the radial coordinates for different values of the NLED parameter and charge $Q$ in Fig. \ref{v_fig}. With increasing radius, the orbital velocity \(v(r)\) of timelike test particles decreases monotonically. It reaches its maximum at relatively nearer regions of the NLED BH and mimics Newtonian \(v\propto r^{-1/2}\) behaviour at large values of \(r\). A larger value of the NLED parameter \(\zeta\) in panel (a) with fixed charge \(Q\) results in a slight upward shift in \(v(r)\) at all radii, suggesting that stronger nonlinear electromagnetic corrections deepen the effective potential well and necessitate higher orbital speeds for circular motion. As an illustration of how the Coulombic effects are related, panel (b) shows that increasing \(Q\) results in a slight increase in \(v(r)\) at fixed \(\zeta\).

\section{Rotating NLED Black Holes}
\label{sec4}
The Newman-Janis Algorithm (NJA) \cite{Newman1965Jun,Newman1965Jun1,Newman1965Jun2} is a powerful, though heuristic, method of generating rotating BH solutions from their non-rotating counterparts. Its popularity derives from the fact that it circumvents the necessity of starting again from scratch in solving the Einstein field equations by following a sophisticated coordinate transformation applied to a static, spherically symmetric metric. Originally it was introduced to extract the rotating Kerr BH solution from the non-rotating Schwarzschild BH metric \cite{Newman1965Jun2,Drake1997Jul}, the NJA has later been extended to several other types of spacetimes \cite{Brauer2015Jan,Lombardo2004Feb,Kim2025Jan,Abbas2024Apr,Alexeyev2025Mar,Jafarzade2025Jun,Abbas2024Apr,Fazzini2025Feb,Li2025Jan,Fathi2025Mar,Zahid2025Feb,Raza2025Jan}. The underlying motivation behind this technique is that most astrophysical BHs are not stationary but rotating, yet nearly all exact BH solutions are stationary because of the simplifications of symmetry needed for solvability.

We shall now proceed to generalize our spherical symmetric NLED BH to a rotating version using the NJA.  To do this, firstly we need to transform the Boyer-Lindquist (BL) coordinates $(t,r,\theta,\phi)$ to the Eddington-Finkelstein (EF) coordinates $(u,r,\theta,\phi)$.  This is done through the following coordinate redefinition\cite{Azreg-Ainou2014Sep,Azreg-Ainou2014Mar,Azreg-Ainou2014May,Jusufi2019Aug}:
	\begin{eqnarray}\label{trans_eq}
		dt&=&du+\frac{dr}{\sqrt{F(r)G(r)}},
	\end{eqnarray}
where the functions $F(r)$ and $G(r)$ arises from the metric:
	\begin{equation}
ds^2 = -F(r)dt^2 + \frac{dr^2}{G(r)} + H(r)(d\theta^2 + \sin^2 \theta d\phi^2).
\label{metric1}
\end{equation}
As per the notations in Ref. \cite{Jusufi2019Aug}, in our framework, we have $f(r) = F(r) = G(r)$ and $H(r) = r^2$.
The metric is conveniently reformulated using null tetrads as
\begin{equation}
	g^{\mu\nu} = -l^{\mu} n^{\nu} - l^{\nu} n^{\mu} + m^{\mu}\overline{m}^{\nu} + m^{\nu}\overline{m}^{\mu},
\end{equation}
where the null tetrads are defined by \cite{Jusufi2019Aug}
\begin{equation}
\begin{aligned}
	l^{\mu} &= \delta^{\mu}_{r},\\[1mm]
	n^{\mu} &= \delta^{\mu}_{u} - \frac{1}{2}F(r)\,\delta^{\mu}_{r},\\[1mm]
	m^{\mu} &= \frac{1}{\sqrt{2H}}\left(\delta^{\mu}_{\theta} + \frac{\dot{\iota}}{\sin\theta}\,\delta^{\mu}_{\phi}\right),\\[1mm]
	\overline{m}^{\mu} &= \frac{1}{\sqrt{2H}}\left(\delta^{\mu}_{\theta} - \frac{\dot{\iota}}{\sin\theta}\,\delta^{\mu}_{\phi}\right).
\end{aligned}
\end{equation}

Notably, the tetrads are chosen so that \(m^\mu\) and \(\overline{m}^\mu\) are complex null vectors, and \(\overline{m}^\mu\) represents the complex conjugate of \(m^\mu\). In the Newman-Penrose formalism, this transformation to complex form is important because it makes it easier to break down spacetime geometry into a null tetrad basis that can better represent the directional properties of the gravitational field.
By definition, the set \(\{l^\mu, n^\mu, m^\mu, \overline{m}^\mu\}\) constitutes a null tetrad that satisfies the orthogonality, normalization and isotropy conditions:
\begin{equation}
\begin{aligned}
	l^{\mu}l_{\mu} = n^{\mu}n_{\mu} = m^{\mu}m_{\mu} = \overline{m}^{\mu}\overline{m}_{\mu} &= 0,\\[1mm]
	l^{\mu}m_{\mu} = l^{\mu}\overline{m}_{\mu} = n^{\mu}m_{\mu} = n^{\mu}\overline{m}_{\mu} &= 0,\\[1mm]
	-l^{\mu}n_{\mu} = m^{\mu}\overline{m}_{\mu} &= 1.
\end{aligned}
\end{equation}
Except for the inner products \(l^\mu n_\mu = -1\) and \(m^\mu \overline{m}_\mu = 1\), which act as normalization conditions, these relations guarantee that all tetrad vectors are null and mutually orthogonal. 

Following the NJA, we can express the coordinate transformation as \cite{Azreg-Ainou2014Sep,Azreg-Ainou2014Mar,Azreg-Ainou2014May,Jusufi2019Aug} 
\begin{equation}
	{x'}^{\mu} = x^{\mu} + ia (\delta_r^{\mu} - \delta_u^{\mu}) \cos\theta,
	\label{NJA1}
\end{equation}
which yields the following relations:
\[
\begin{cases}
	u' = u - ia\cos\theta, \\
	r' = r + ia\cos\theta, \\
	\theta' = \theta, \\
	\phi' = \phi,
\end{cases}
\]
where \(a\) represents the spin parameter. Under this transformation, the null tetrad vectors \(Z^\alpha\) transform according to
\(
	Z^\mu = \frac{\partial x^\mu}{\partial {x^\prime}^\nu}\, {Z^\prime}^\nu,
\)
that results into \cite{Azreg-Ainou2014Sep,Azreg-Ainou2014Mar,Azreg-Ainou2014May,Jusufi2019Aug}:
\begin{equation}
\begin{aligned}
	l'^{\mu} &= \delta^{\mu}_{r},\\[1mm]
	n'^{\mu} &= \sqrt{\frac{B}{A}}\delta^{\mu}_{u} - \frac{1}{2}B\delta^{\mu}_{r},\\[1mm]
	m'^{\mu} &= \frac{1}{\sqrt{2\,\Sigma}}\left[(\delta^{\mu}_{u}-\delta^{\mu}_{r})\dot{\iota}\,a\sin\theta + \delta^{\mu}_{\theta} + \frac{\dot{\iota}}{\sin\theta}\,\delta^{\mu}_{\phi}\right],\\[1mm]
	\overline{m}'^{\mu} &= \frac{1}{\sqrt{2\,\Sigma}}\left[(\delta^{\mu}_{u}-\delta^{\mu}_{r})\dot{\iota}\,a\sin\theta + \delta^{\mu}_{\theta} + \frac{\dot{\iota}}{\sin\theta}\,\delta^{\mu}_{\phi}\right].
\end{aligned}
\label{nj_eqs}
\end{equation}
Here it is assumed that the functions \((G(r), F(r), H(r))\) transform to \((A(a,r,\theta), B(a,r,\theta), \Sigma(a,r,\theta))\). This notion was first proposed by Azreg-A\"{i}nou in Ref. \cite{Azreg-Ainou2014Sep}, where he developed a non-complexification procedure to generate rotating BH and non-BH solutions in GR. He showed that this technique is applicable to study the properties of Ay\'on--Beato--Garc\'ia BH, which belongs to the class of NLED BHs. The technique is further extended to study wormholes with imperfect fluids with and without electric and magnetic fields \cite{Azreg-Ainou2014May} and in generating some general solutions of an imperfect fluid and its conformal homologous counterpart \cite{Azreg-Ainou2014Mar}. In Ref. \cite{Azreg-Ainou2014Mar}, the author stressed that the use of conformal fluids as cores of both static and rotating fluids does not suffer from malicious behaviour in terms of regularity of the solutions.

With the null tetrad vectors defined, the contravariant components of the new metric can be constructed as:
\begin{eqnarray}
\begin{aligned}
	g^{uu} &= \frac{a^{2}\sin^{2}\theta}{\Sigma},  & g^{u\phi} &= \frac{a}{\Sigma},                  & g^{ur} &= 1-\frac{a^{2}\sin^2\theta}{\Sigma}, \\[1mm]
	g^{rr} &= \mathcal{F}+\frac{a^{2}\sin^{2}\theta}{\Sigma}, & g^{r\phi} &= -\frac{a}{\Sigma},  & g^{\theta\theta} &= \frac{1}{\Sigma}, \\[1mm]
	g^{\phi\phi} &= \frac{1}{\Sigma\sin^2\theta}.
\end{aligned}
\label{metric_terms}
\end{eqnarray}

Here, \(\Sigma = r^2 + a^2 \cos^2\theta\) and \(\mathcal{F}\) is a function of \(r\) and \(\theta\). The metric is then given by
\begin{align}
	ds^2 ={} & -\mathcal{F}\,du^2 - 2\,du\,dr + 2a\sin^2\theta\left(\mathcal{F}-1\right)du\,d\phi \nonumber\\[1mm]
	&+ 2a\sin^2\theta\,dr\,d\phi + \Sigma\,d\theta^2 \nonumber\\[1mm]
	&+ \sin^2\theta\left[\Sigma + a^2\left(2-\mathcal{F}\right)\sin^2\theta\right]d\phi^2.
\end{align}
Furthermore, the black hole solution can be reexpressed in the original coordinates by applying the transformations
\begin{equation}
	du = dt - \frac{a^2+r^2}{\Delta}\,dr,\quad d\phi = d\varphi - \frac{a}{\Delta}\,dr,
\end{equation}
with \(\Delta\) defined as \cite{Azreg-Ainou2014Sep,Azreg-Ainou2014Mar,Azreg-Ainou2014May,Jusufi2019Aug}
\begin{equation}
\begin{aligned}
	\Delta &= r^{2}f(r) + a^{2} \\&= a^2 - 2 M r + r^2 - \frac{4}{3}\zeta \sqrt{2Q^3}\,\, \log \left(2 \sqrt{2 Q}\,\zeta \right) r^2,
	\end{aligned}
\end{equation}
where $f(r)$ is the lapse function of the NLED BH assumed in our work. Notice that as, $\zeta = Q = 0$, the usual Kerr solution $\Delta = a^2 - 2 M r + r^2 $  is obtained.
	
	\subsection{Shape of Ergoregion}
\begin{figure*}[htb]
\includegraphics[scale=0.5]{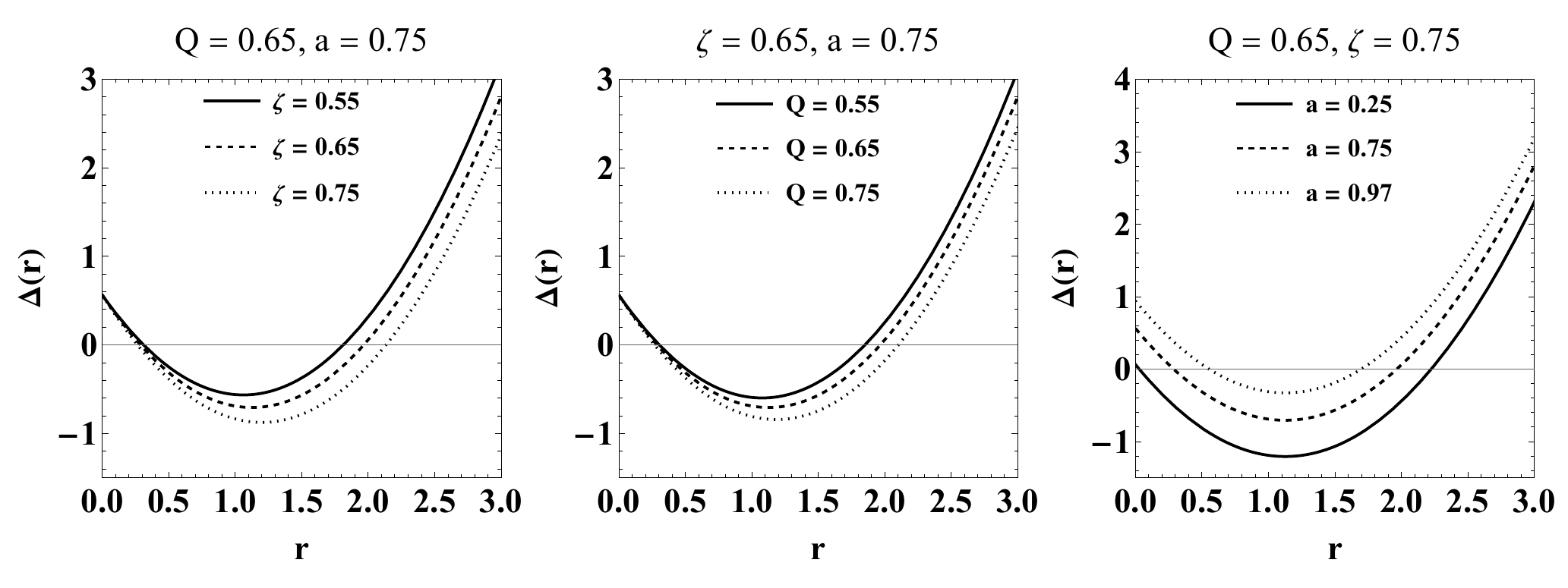}
\caption{The plot of the function $\Delta(r)$ is shown for various values of the NLED parameter, charge $Q$ and the spin parameter $a$.}
\label{delta_plot}
\end{figure*}

	Let us now proceed to study the shape of the ergoregion of our black hole metric given by \eqref{lapse1}. In particular, we shall be interested in plotting the shape of the ergoregion in the $xz$-plane. Recall that the horizons of the black hole can be found by solving $\Delta=0$, i.e., 
\begin{equation}
a^2 - 2 M r + r^2 - \frac{4}{3}\zeta \sqrt{2Q^3}\,\, \log \left(2 \sqrt{2 Q}\,\zeta \right) r^2 =0,
\end{equation}
The horizons are obtained at
\begin{widetext}
\begin{equation}
\begin{aligned}
r_{-} &= \frac{1}{3} \left(-\sqrt{\left(3 M+2 \sqrt{2} \zeta  Q^{3/2} \log \left(2 \sqrt{2} \zeta  \sqrt{Q}\right)\right)^2-9 a^2}+3 M+2 \sqrt{2} \zeta  Q^{3/2} \log \left(2 \sqrt{2} \zeta  \sqrt{Q}\right)\right), \\
r_{+} &= \frac{1}{3} \left(\sqrt{\left(3 M+2 \sqrt{2} \zeta  Q^{3/2} \log \left(2 \sqrt{2} \zeta  \sqrt{Q}\right)\right)^2-9 a^2}+3 M+2 \sqrt{2} \zeta  Q^{3/2} \log \left(2 \sqrt{2} \zeta  \sqrt{Q}\right)\right).
\end{aligned}
\end{equation}
\end{widetext}
The static limit region, also known as the inner and outer ergosurface, can be obtained  by $g_{tt}=0$, i.e., 
	\begin{equation}
a^2 \cos^2 \theta - 2 M r + r^2 - \frac{4}{3}\zeta \sqrt{2Q^3}\log \left(2 \sqrt{2 Q}\,\zeta \right) r^2 =0,
\end{equation}
The plot showing the event horizons is given in Fig. \ref{delta_plot}

\begin{figure*}[htb]
\centerline{
\includegraphics[scale=0.3]{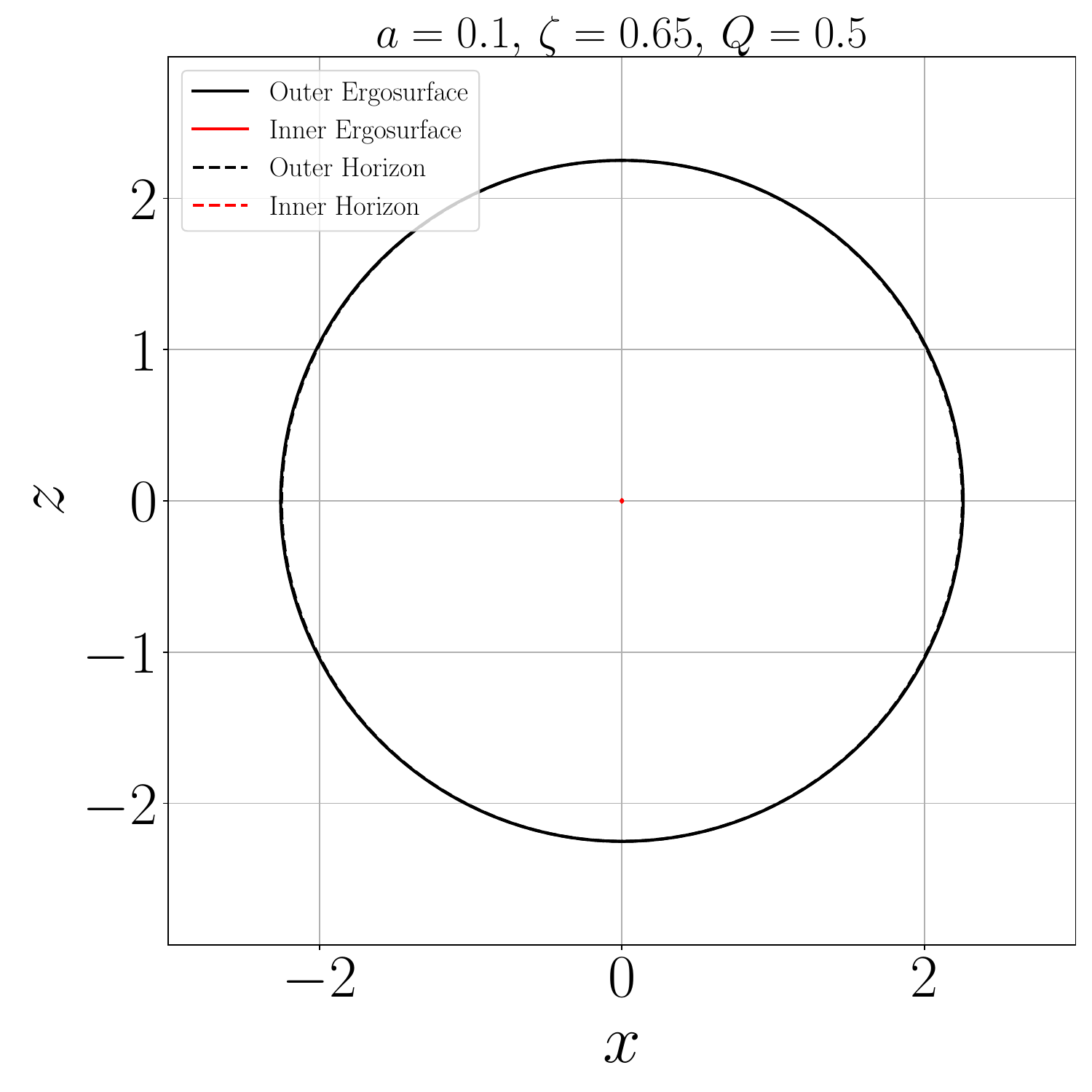}
\includegraphics[scale=0.3]{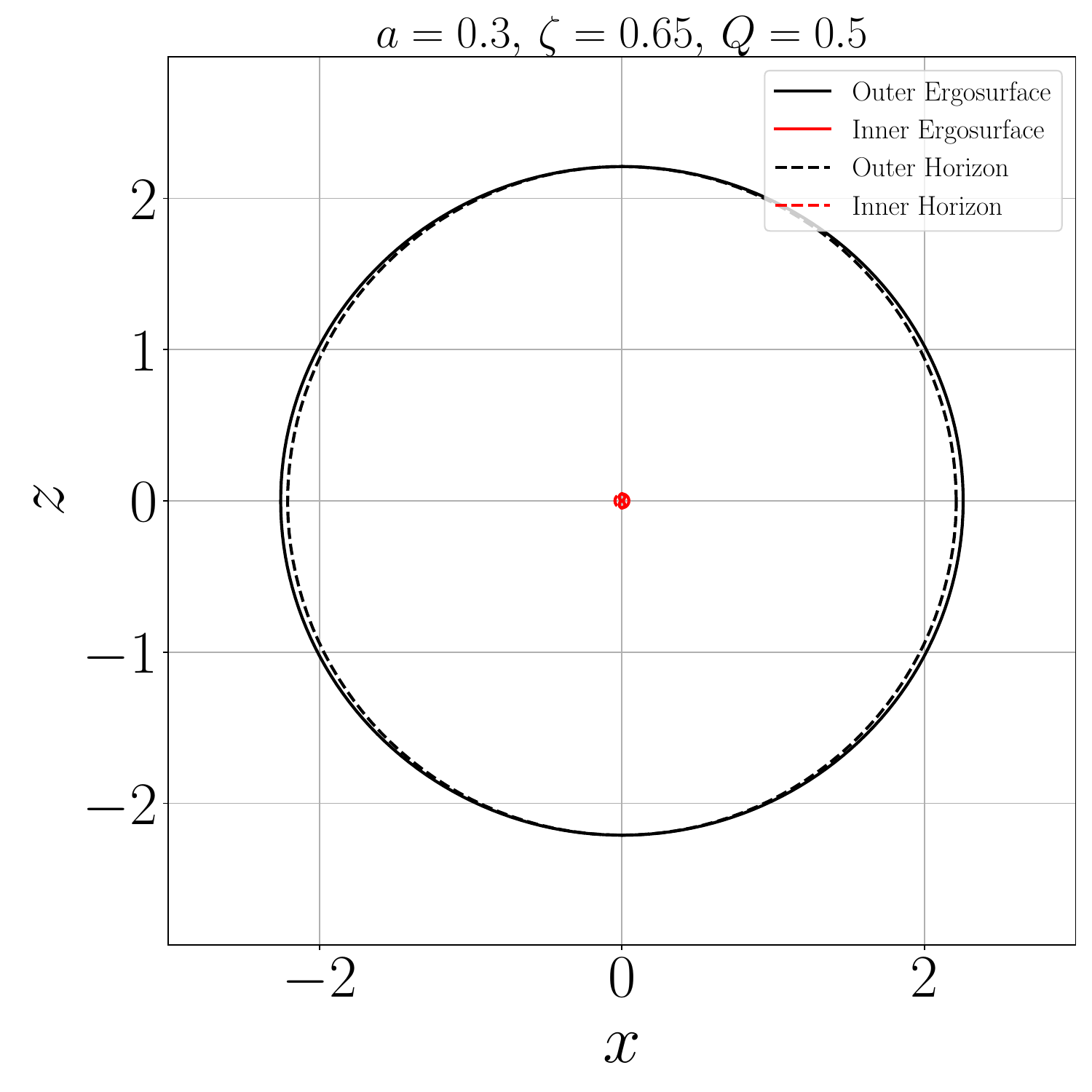}}
\centerline{
\includegraphics[scale=0.3]{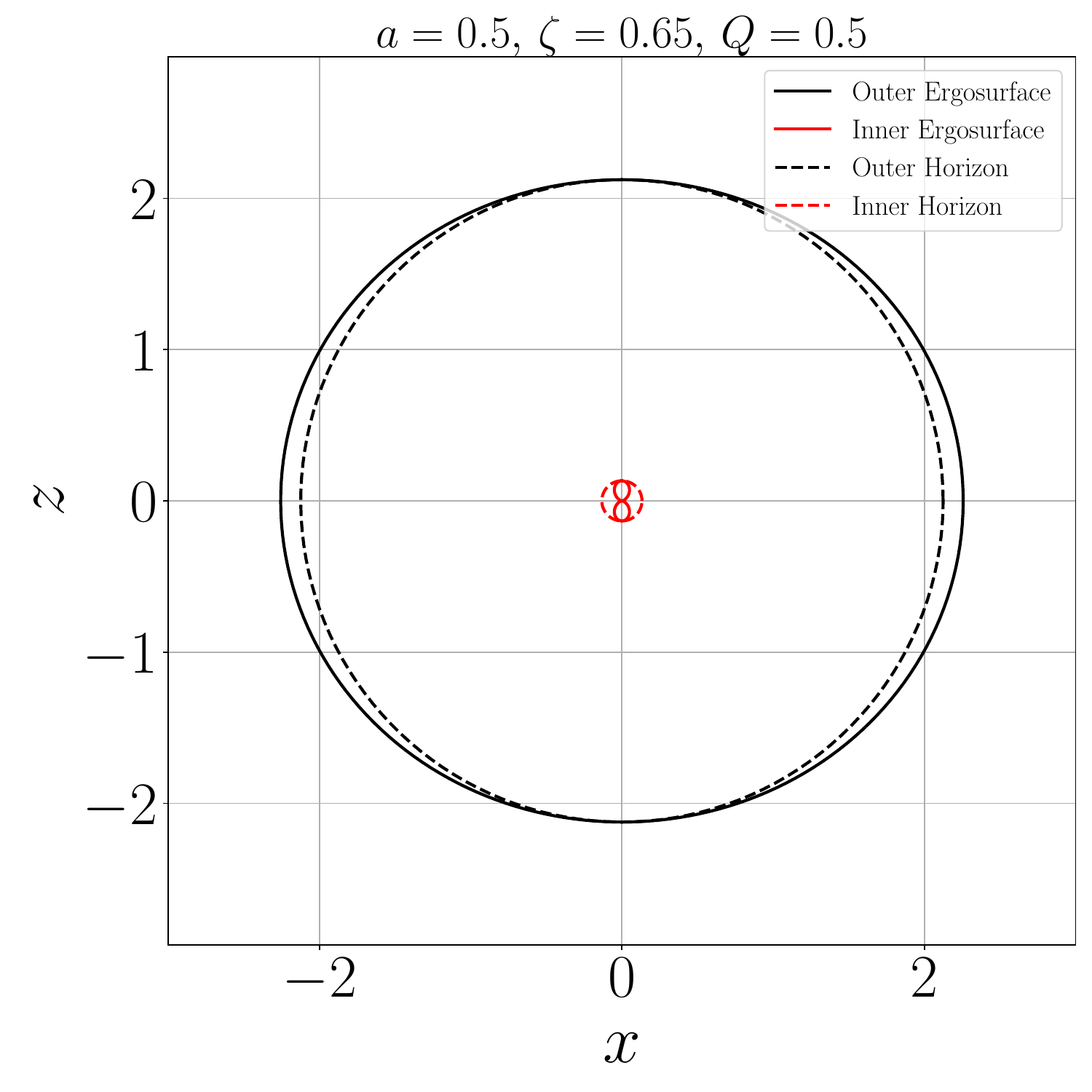}
\includegraphics[scale=0.3]{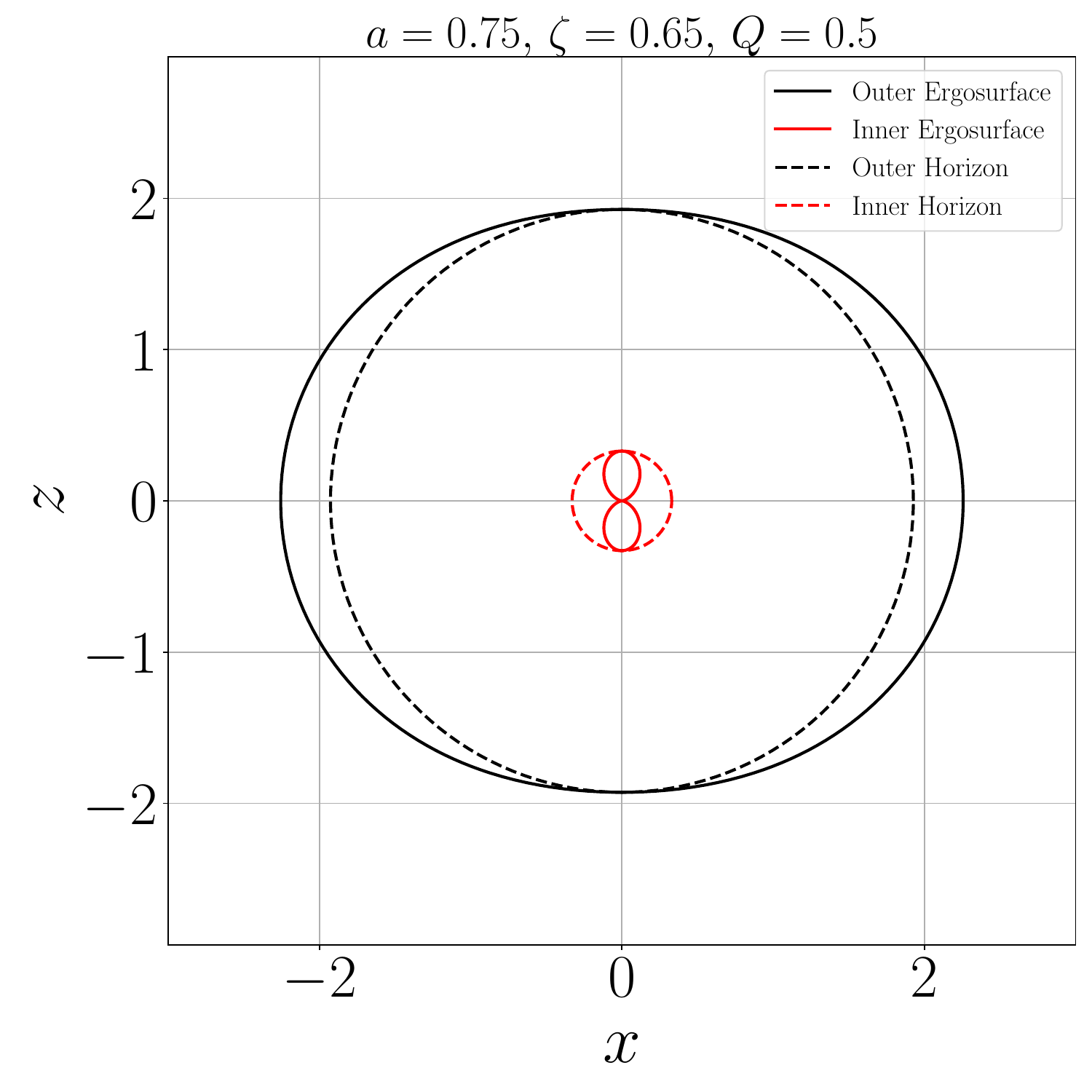}}
\centerline{
\includegraphics[scale=0.3]{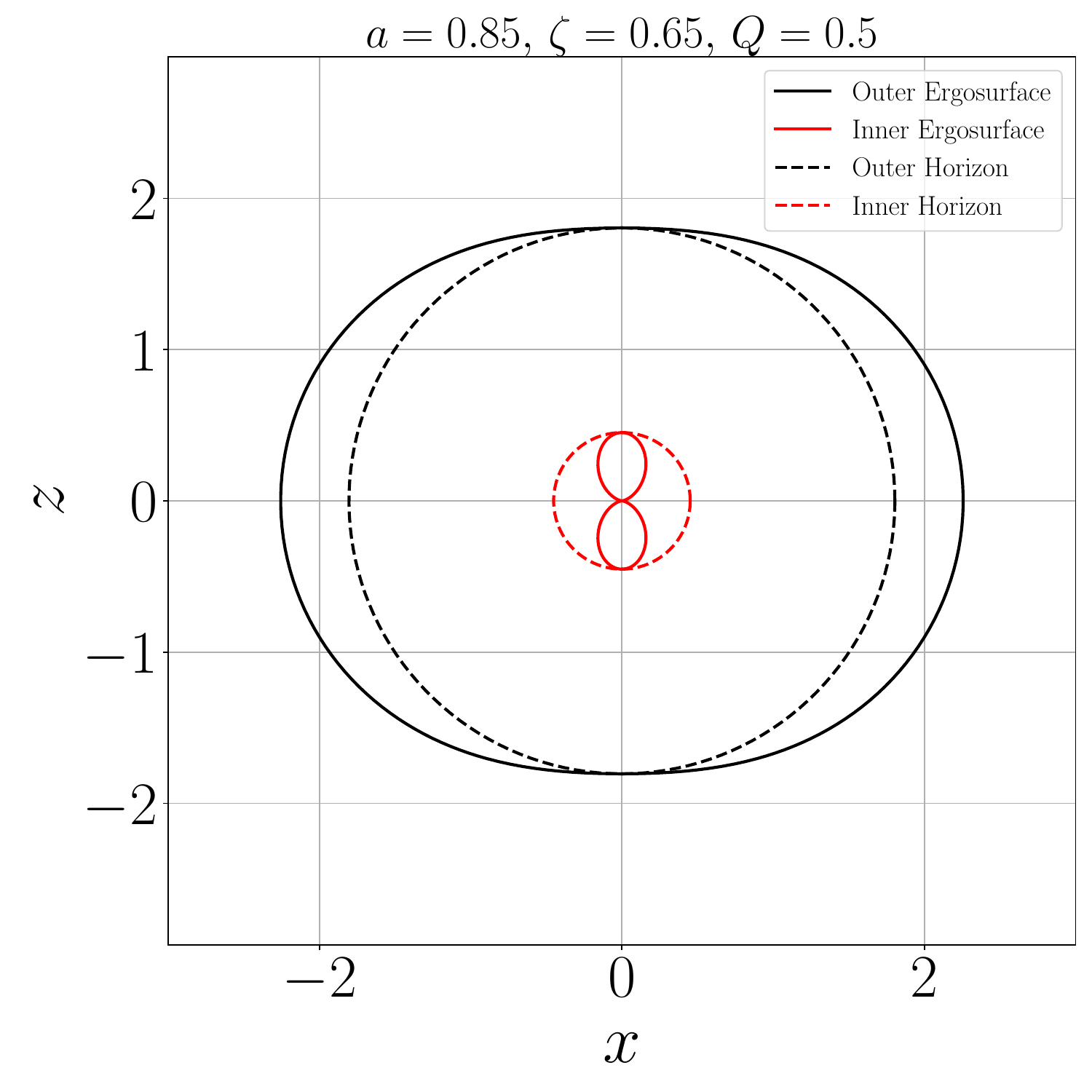}
\includegraphics[scale=0.3]{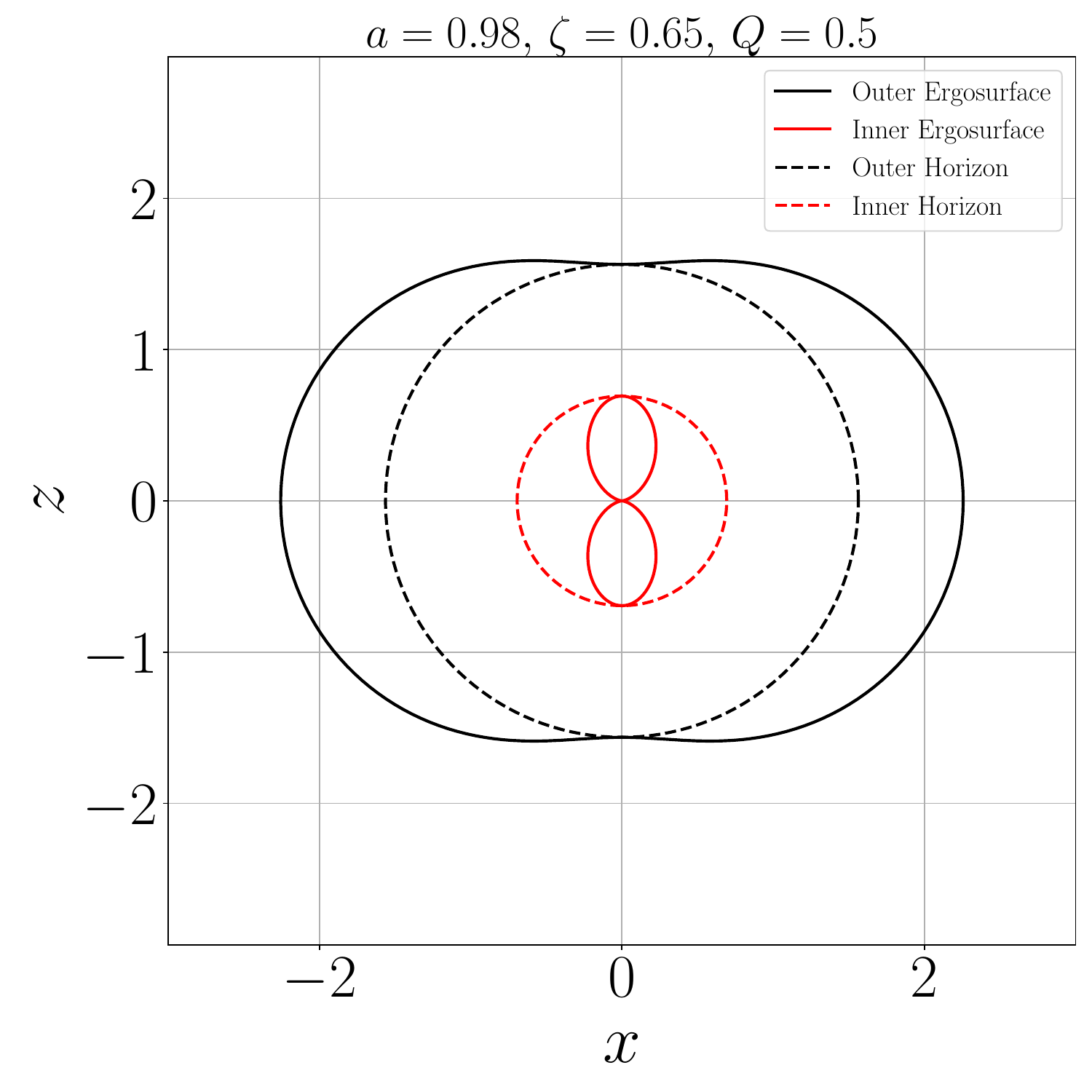}}
\caption{The 2D ergo region is plotted for various values of spin parameter $a$.}
\label{ergo_fig_2d}
\end{figure*}

\begin{figure*}[htb]
\centerline{
\includegraphics[scale=0.25]{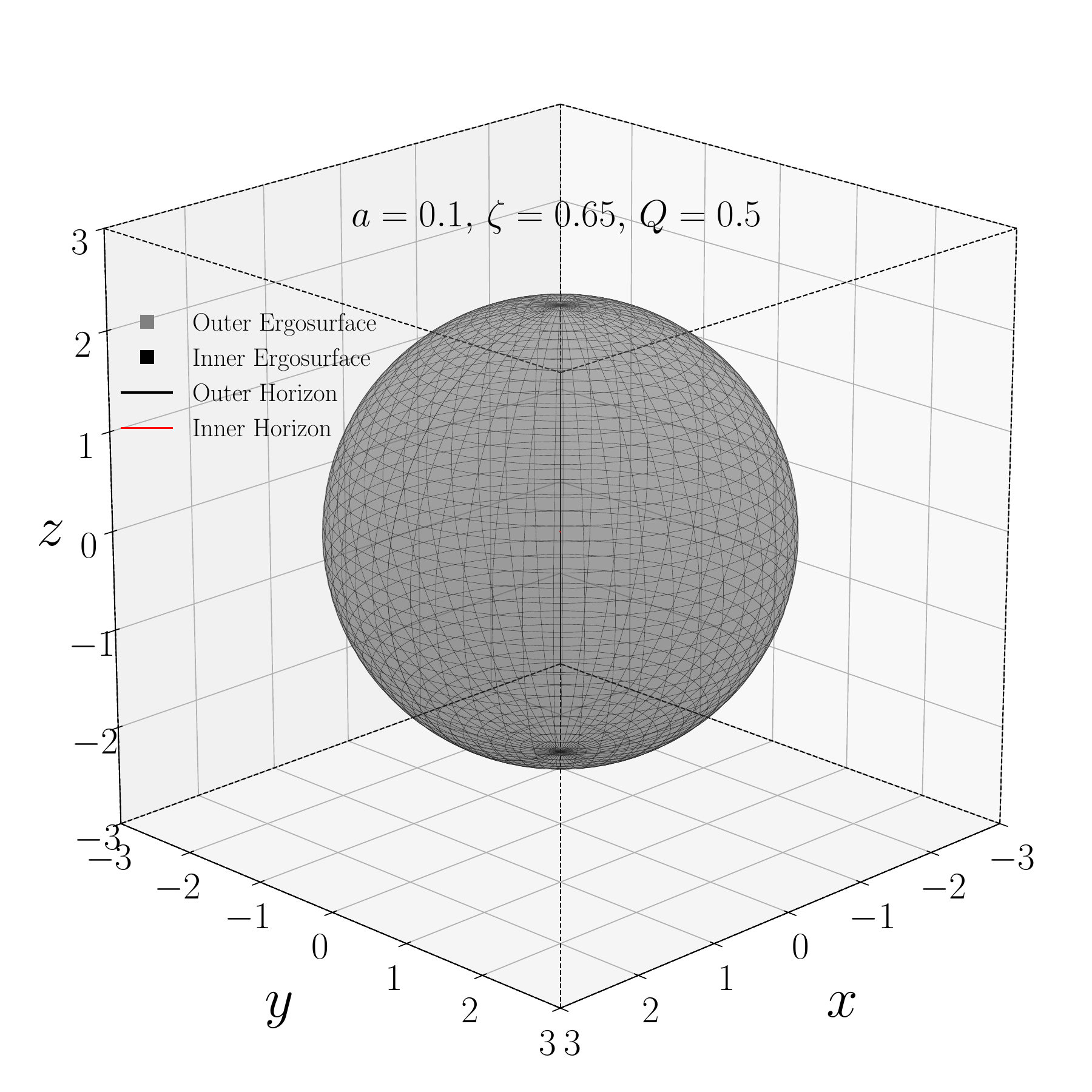}
\includegraphics[scale=0.25]{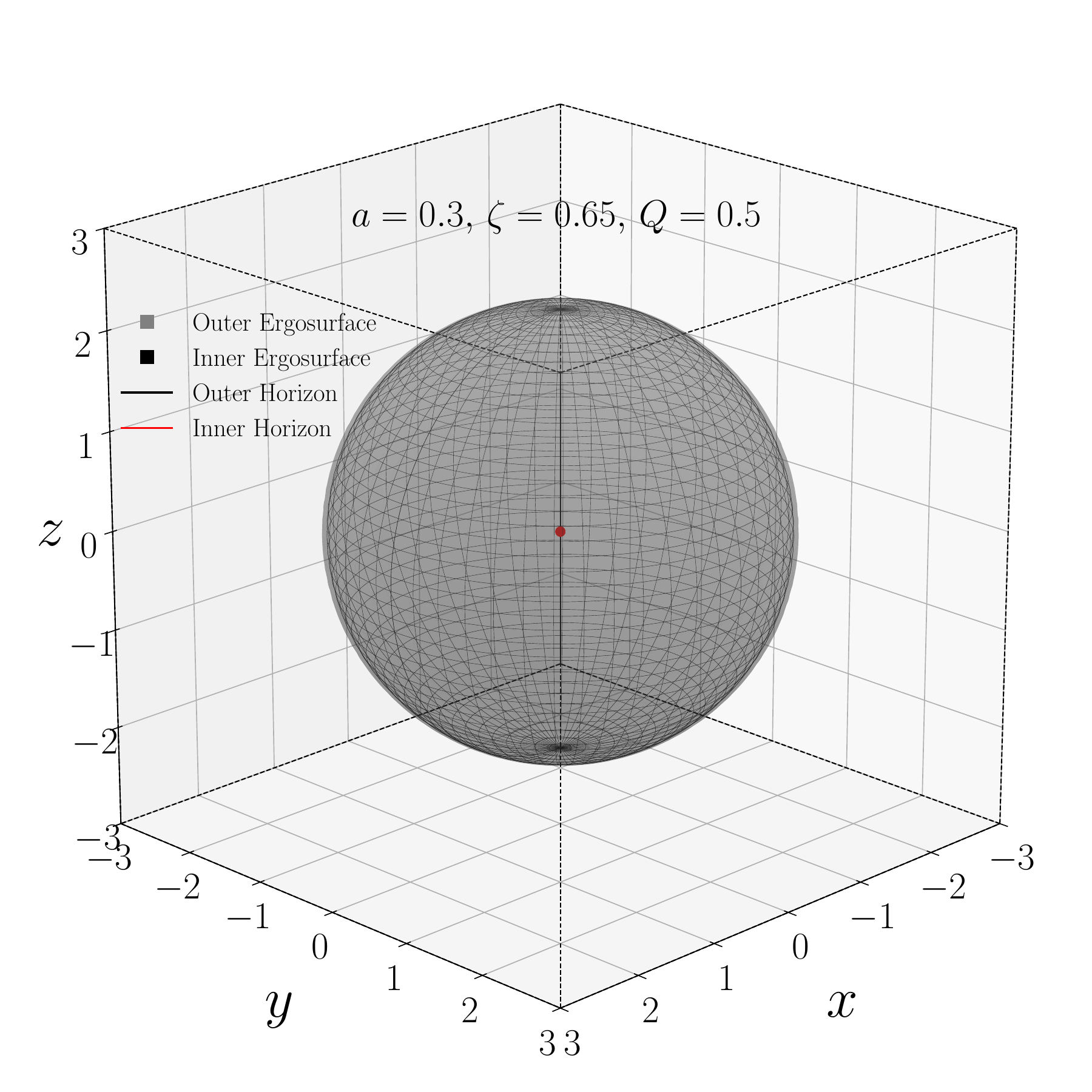}}
\centerline{
\includegraphics[scale=0.25]{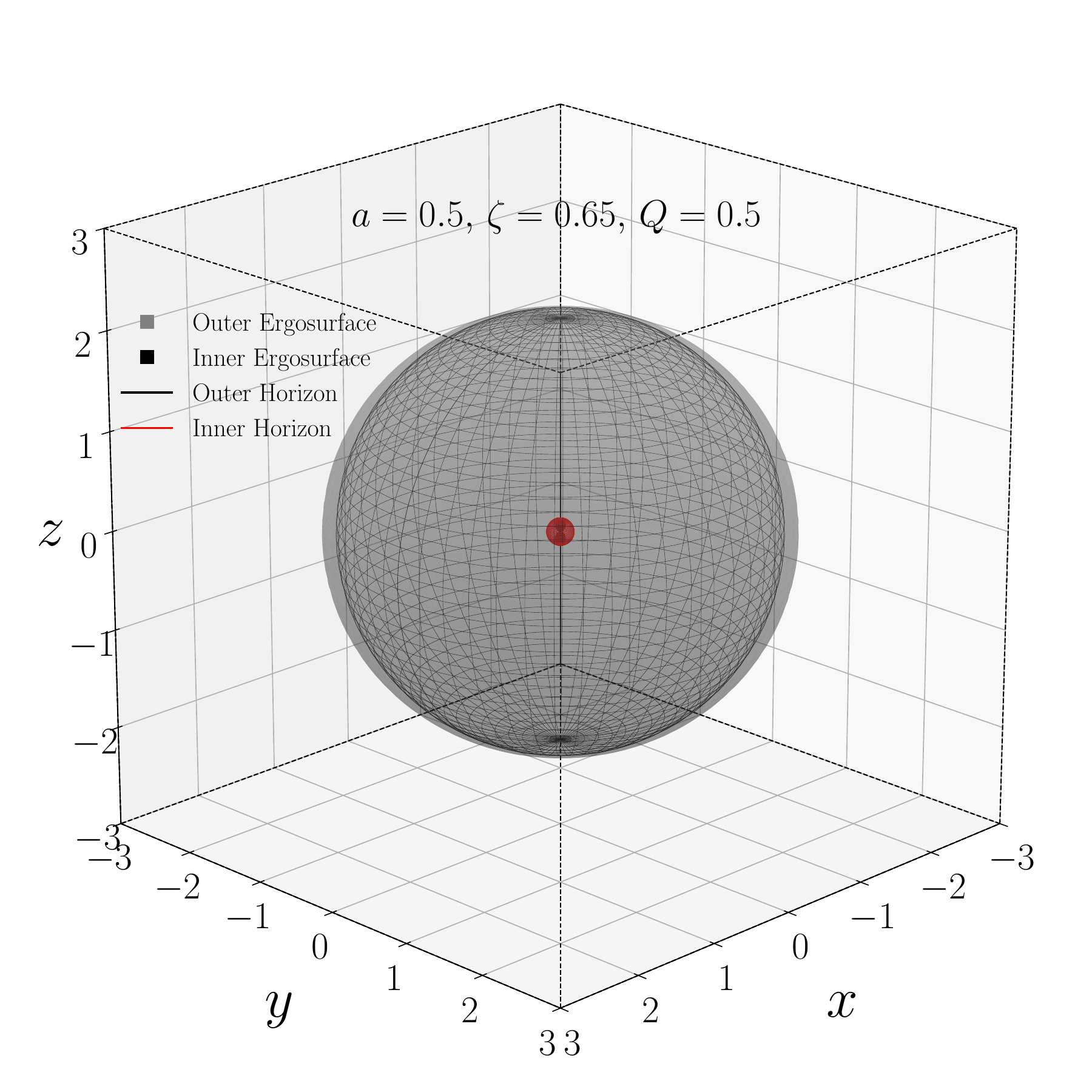}
\includegraphics[scale=0.25]{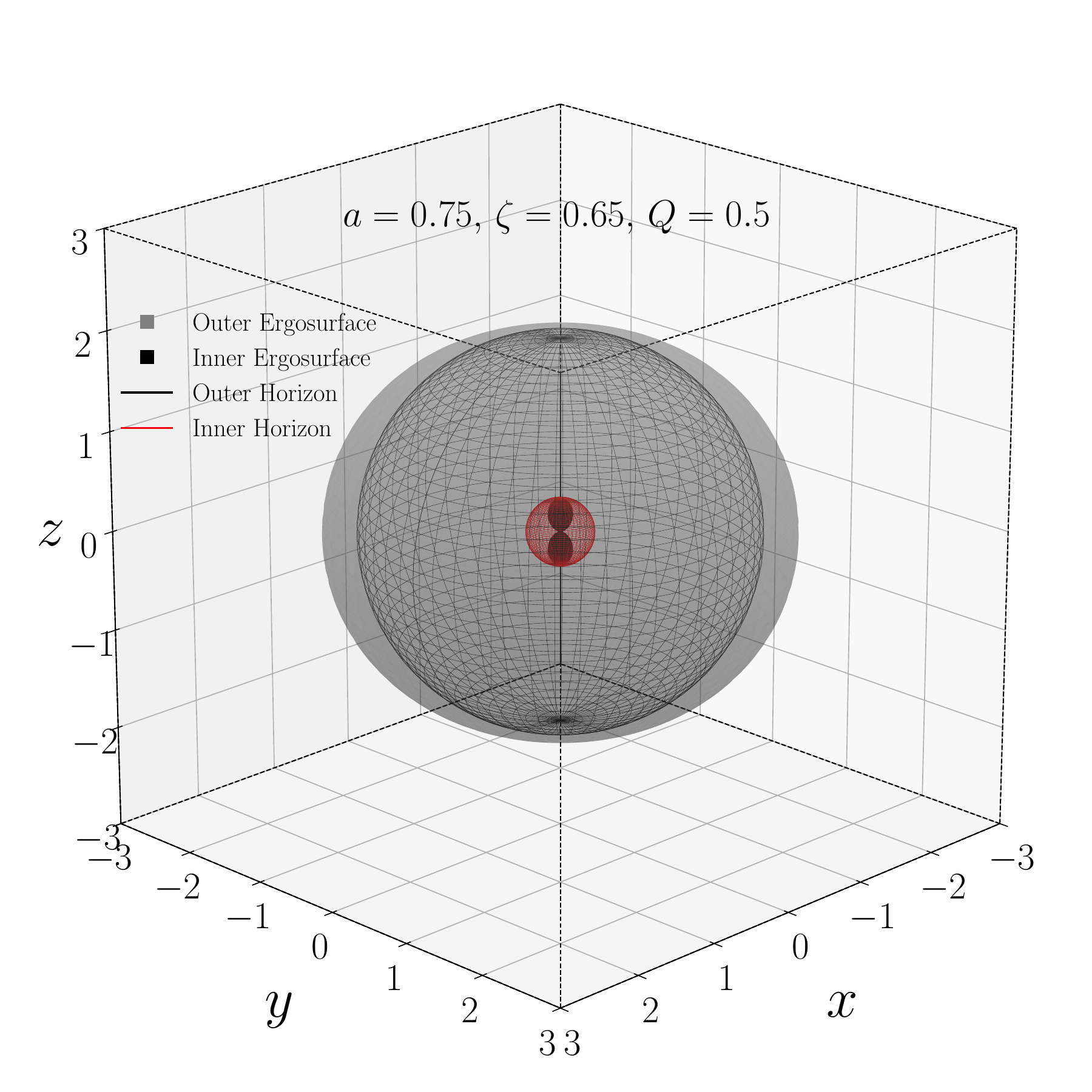}}
\centerline{
\includegraphics[scale=0.25]{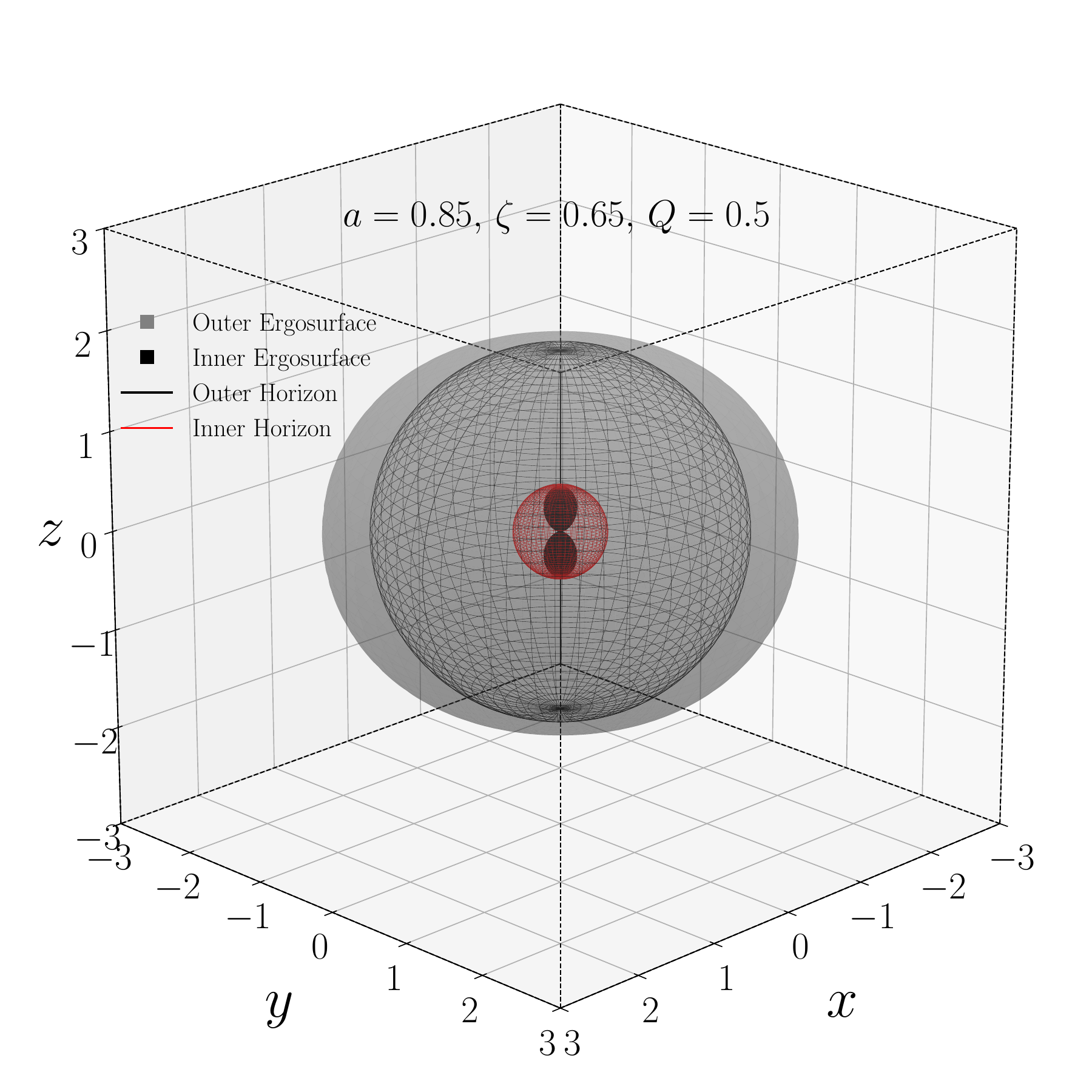}
\includegraphics[scale=0.25]{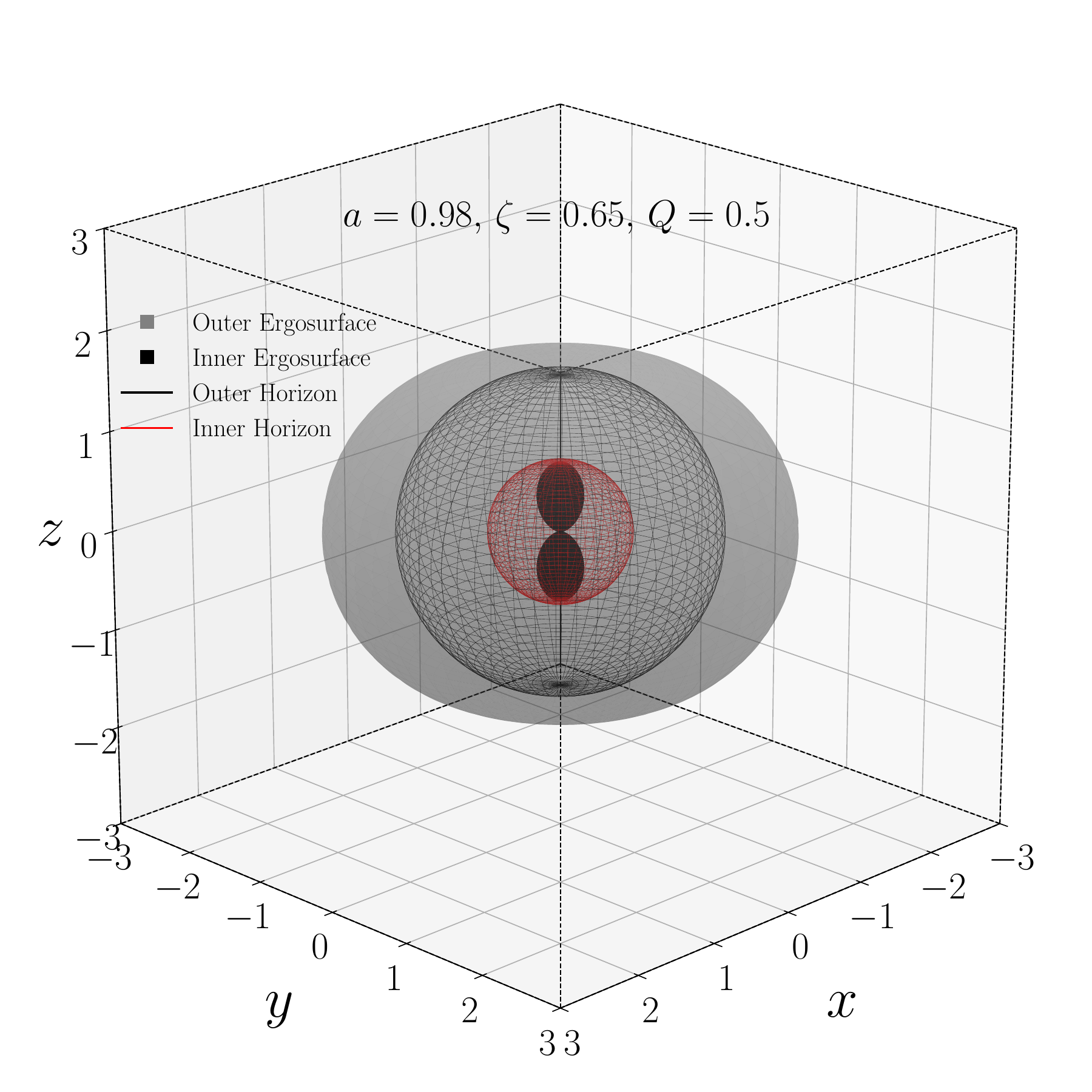}}
\caption{The 3D ergo region is plotted for various values of spin parameter $a$.}
\label{ergo_fig_3d}
\end{figure*}

The ergosurfaces and horizons are displayed in the figures \ref{ergo_fig_2d} and \ref{ergo_fig_3d}.

	\section{The Hamilton Jacobi Equation: Shadow}
	\label{sec5}
In this section, let us determine the shadow cast by the rotating NLED BH. To do that, we first construct the geodesic equations of motion of photons. To obtain the null geodesics, we follow the Hamilton-Jacobi formalism as follows: 
	\begin{equation}
		\partial_\tau\mathcal{J}=-\mathcal{H}.
		%\frac{\partial S}{\partial \tau}=-\frac{1}{2}g^{\mu\nu}\frac{\partial S}{\partial {x^{\mu}}}\frac{\partial S}{\partial {x^{\nu}}}
		\label{HJ1}
	\end{equation}
	Here, $\mathcal{J}$ denotes the Jacobi action, which can be defined in terms of the affine parameter $\tau$ and the coordinates $x^\mu$ as $\mathcal{J}=\mathcal{J}(\tau,x^\mu)$ and $\mathcal{H}$ denotes the Hamiltonian of the particle, as given by $g^{\mu\nu}\partial_\mu\mathcal{J}\;\partial_\nu\mathcal{J}$. The spacetime symmetries dictate that photon energy $E$ and angular momentum $L$ must be conserved quantities, which are given by the usual definition of Killing fields $\kappa_t=\partial_t$ and $\kappa_\phi=\partial_\phi$, respectively. It is well known that the Hamilton-Jacobi equation can be solved in terms of separable solutions, which contain the already existing conserved quantities, i.e,
	\begin{equation}
		\mathcal{J}=\frac{1}{2}m^2\tau-Et+L\phi+\mathcal{J}_r(r)+\mathcal{J}_{\theta}(\theta)
		\label{HJ2}
	\end{equation}
	with $\mathcal{J}_r(r)$ and $\mathcal{J}_\theta(\theta)$ are functions of the coordinates $r$ and $\theta$.

	Combining Eq. (\ref{HJ1}) and Eq. (\ref{HJ2}), one can obtain the geodesic equations of motion expressed as the four velocity components as follows:
		\begin{align}
	\label{HJ3}
	&\Sigma\frac{dt}{d\tau}=\frac{r^2+a^2}{\Delta}[E(r^2+a^2)-aL]-a(aE\sin^2\theta-L),\\
	&\Sigma\frac{dr}{d\tau}=\sqrt{\mathcal{R}(r)},\\
	&\Sigma\frac{d\theta}{d\tau}=\sqrt{\Theta(\theta)},\\
	\label{HJ4}
	&\Sigma\frac{d\varphi}{d\tau}=\frac{a}{\Delta}[E(r^2+a^2)-aL]-\left(aE-\frac{L}{\sin^2\theta}\right),
	\end{align}
	where $\mathcal{R}(r)$ and $\Theta(\theta)$ are expressed by
	\begin{align}
	\label{HJ5}
	&\mathcal{R}(r)=[E(r^2+a^2)-aL]^2-\Delta[m^2r^2+(aE-L)^2+\mathcal{K}],\\
	&\Theta(\theta)=\mathcal{K}-\left(  \dfrac{L^2}{\sin^2\theta}-a^2E^2  \right) \cos^2\theta.
	\end{align}
	Here. $\mathcal{K}$ is a separation constant known as the Carter constant. This is an additional constant of motion arising from a hidden symmetry in rotating spacetimes that ensures separability and full integrability of the geodesic equations of motion.

Depending on the value of the impact parameter, the photons from a light source may eventually be scattered away or fully captured by the black hole. However, for some critical value of the impact parameter, the photon may result in a photon sphere.  This behavior shows the area that delineates the edge of the shadow.  Using the effective potential, \(V_{\text{eff}}\), related to the radial motion of the photon, to define the radial geodesic equation, we may investigate the existence of unstable circular orbits around the black hole, which is defined by:
	\begin{equation}
	\Sigma^2\left(\frac{dr}{d\tau}\right)^2+V_{\text{eff}}=0.
	\end{equation}
	
Let us define two parameters $\zeta$ and $\eta$ \cite{Perlick2022Feb} as:
	\begin{equation}
	\xi=L/E,  \quad \text{and} \quad   \eta=\mathcal{K}/E^2.
	\end{equation}
	In terms of these two parameters, the effective potential can be redefined as:
	\begin{equation}\label{veff}
	V_{\text{eff}}=\Delta((a-\xi)^2+\eta)-(r^2+a^2-a\;\xi)^2.
	\end{equation}
	Here, we can safely replace $V_{\text{eff}}/E^2$ by $V_{\text{eff}}$ for convenience.\footnote{Note that this replacement does not affect the position of critical orbits. It merely shifts the peak of the effective potential towards a higher value, while keeping the position of the photon orbit intact.}. The critical photon orbits corresponding to a constant radius $r=r_c$ satisfy the following conditions:
	\begin{equation}\label{cond}
	V_{\text{eff}}(r)=0,\quad~~~\frac{dV_{\text{eff}}(r)}{dr}=0
	\end{equation}
	Using the conditions \eqref{cond}, we can obtain the parameters $\xi$ and $\eta$ for the rotating analogue of the NLED BH as:
	\begin{widetext}

It follows that $\xi$ and $\eta$ are directly related to the function $\Delta$ through the relations 
\begin{equation}
\xi(r) = \frac{\left(a^2+r^2\right) \Delta '(r)-4 r \Delta (r)}{a \Delta '(r)},
\end{equation}	
	and
	\begin{equation}
	\eta (r) = \frac{16 r^2 \Delta (r) \left(a^2-\Delta (r)\right)+r^4 \left(-\Delta '(r)^2\right)+8 r^3 \Delta (r) \Delta '(r)}{a^2 \Delta '(r)^2}.
	\end{equation}
	Thus, we have for our system
	\begin{equation}
	\xi(r) = \frac{3 a^2 (M+r)+2 \sqrt{2} \zeta  Q^{3/2} \left(a^2-3 r^2\right) \log \left(2 \sqrt{2} \zeta  \sqrt{Q}\right)+3 r^2 (r-3 M)}{a \left(3 M+2 \sqrt{2} \zeta  Q^{3/2} \log \left(2 \sqrt{2} \zeta  \sqrt{Q}\right)-3 r\right)}
	\label{xi_eq}
	\end{equation}
	
	\begin{equation}
\eta(r) = \frac{3 r^3 \left(4 \sqrt{2} \zeta  Q^{3/2} \left(2 a^2+3 r (r-3 M)\right) \log \left(2 \sqrt{2} \zeta  \sqrt{Q}\right)+3 \left(4 a^2 M-r (r-3 M)^2\right)-24 \zeta ^2 Q^3 r \log ^2\left(2 \sqrt{2} \zeta  \sqrt{Q}\right)\right)}{a^2 \left(3 M+2 \sqrt{2} \zeta  Q^{3/2} \log \left(2 \sqrt{2} \zeta  \sqrt{Q}\right)-3 r\right)^2}
	\end{equation}
	\end{widetext}
	To plot the shadow images, we need to use the coordinates $\alpha$ and $\beta$, which are defined through 
	
	\begin{equation}
		\begin{aligned}
		\alpha &= \lim_{r_0\to \infty} \left(-r_0^2 \sin \theta_0 \frac{d\phi}{dr} \Big|_{(r_0, \theta_0)}\right), \\ \beta &= \lim_{r_0\to \infty} \left(r_0^2 \frac{d\theta}{dr}\Big|_{(r_0, \theta_0)}\right).
		\end{aligned}
		\label{xy_shad}
	\end{equation}
	with $(r_0, \theta_0)$ denoting the observer's position.
	
	For any asymptotically flat spacetime, we take the $r \to \infty$ limit, which gives
	\begin{equation}
		\begin{aligned}
		\alpha &= - \frac{\xi}{\sin \theta_0}, \\
		\beta &= \pm \sqrt{\eta + a^2 \cos^2 \theta_0 - \xi^2 \cot^2 \theta_0}
		\end{aligned}
		\label{xy_shad2}
	\end{equation}
	
	When the observer is situated at the equatorial plane, one can set $\theta_0 = \pi/2$, which reduces Eq. \eqref{xy_shad2} to 
	\begin{equation}
			\begin{aligned}
			\alpha &= - \xi, \\
			\beta &= \pm \sqrt{\eta}
		\end{aligned}
		\label{xy_shad_eq}
	\end{equation}
		
\begin{figure*}[htbp]
\centerline{\includegraphics[scale=0.35]{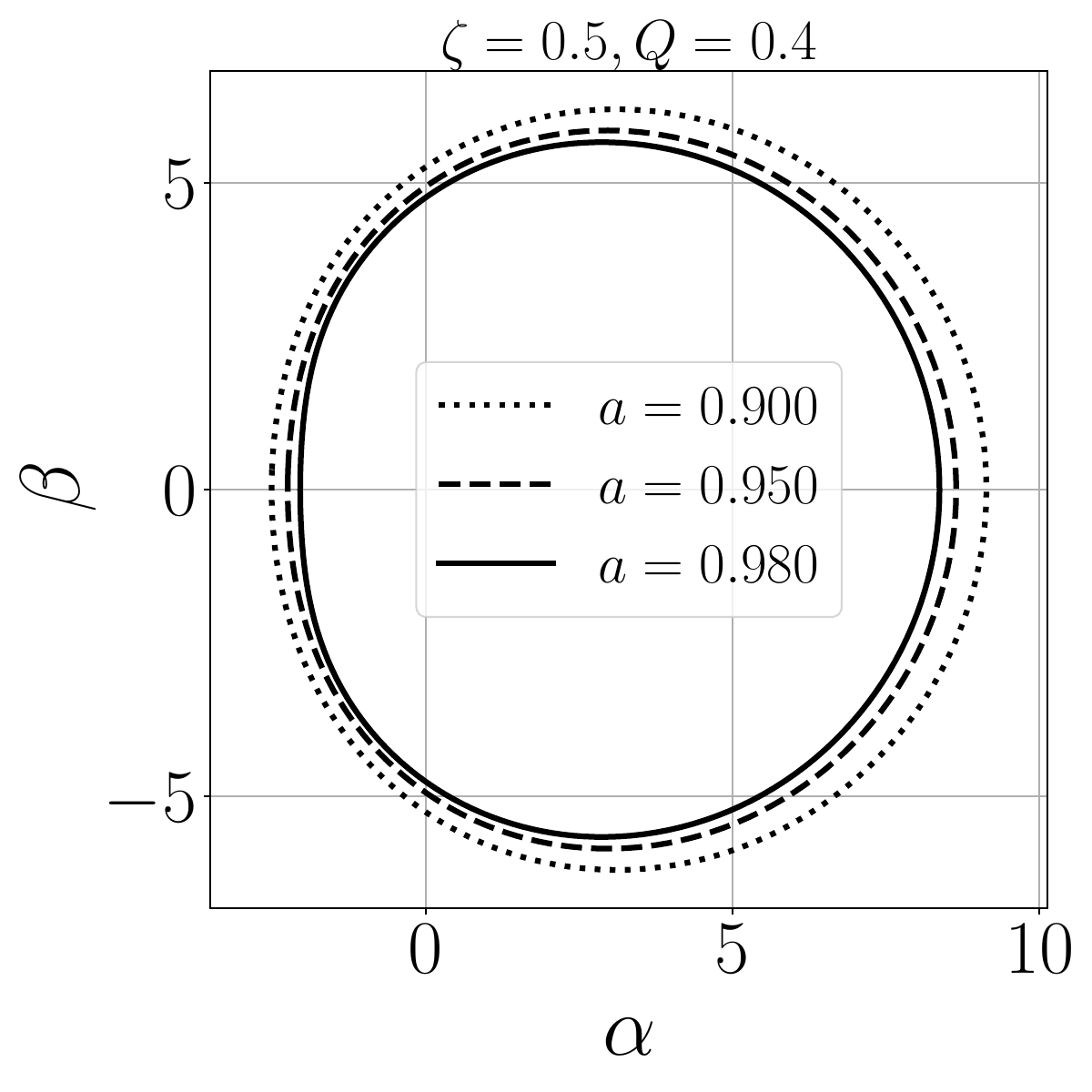}\includegraphics[scale=0.35]{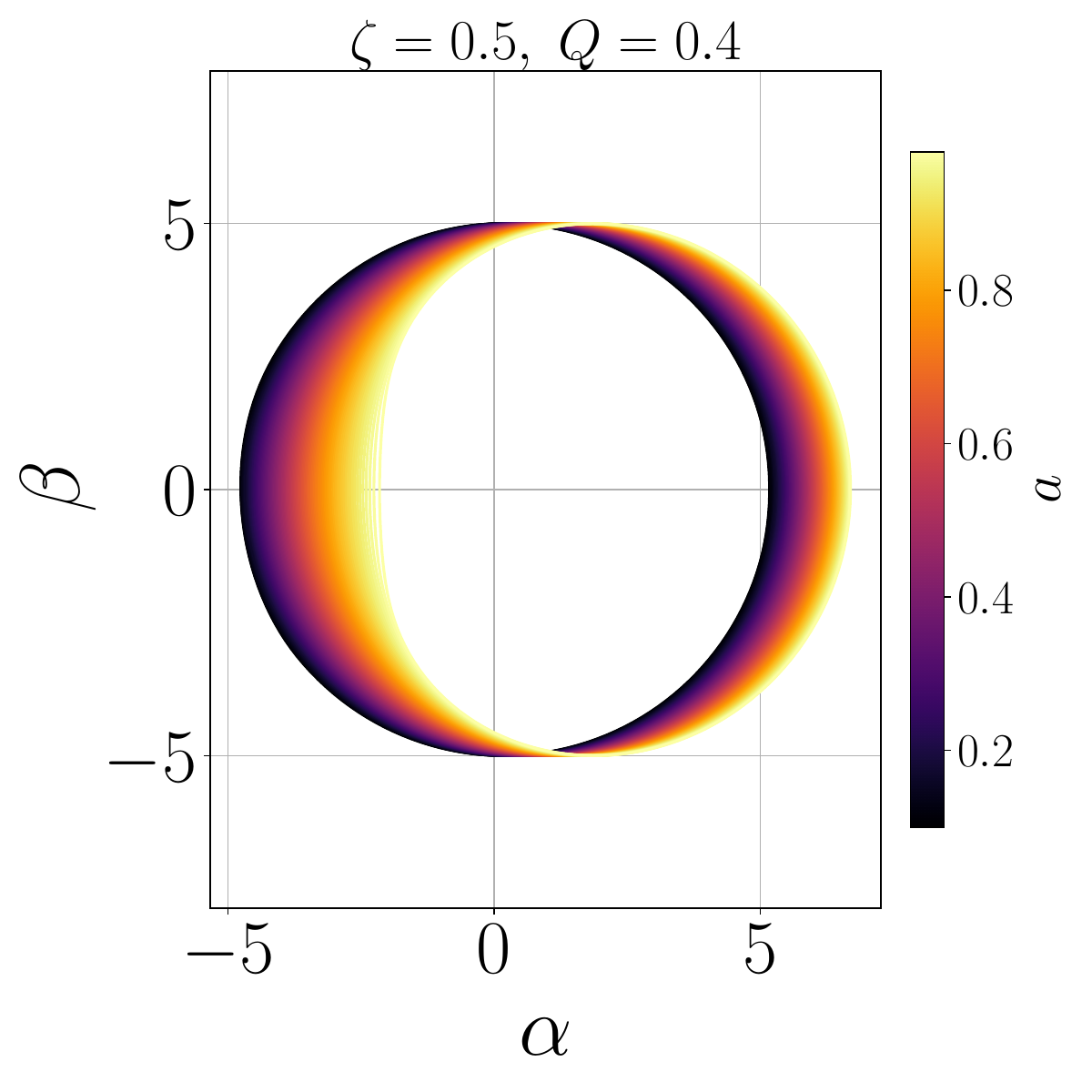}}
\centerline{\includegraphics[scale=0.35]{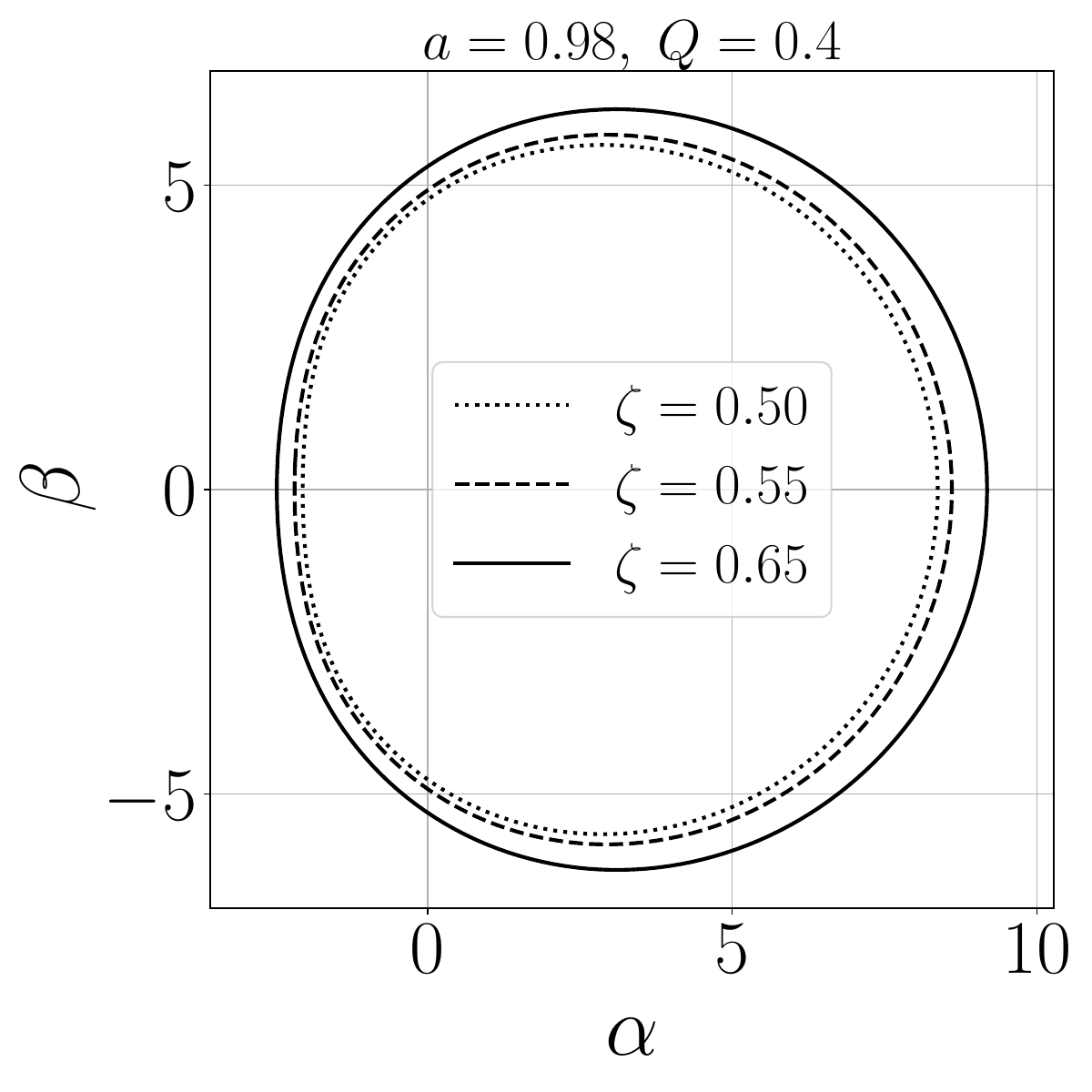} \includegraphics[scale=0.35]{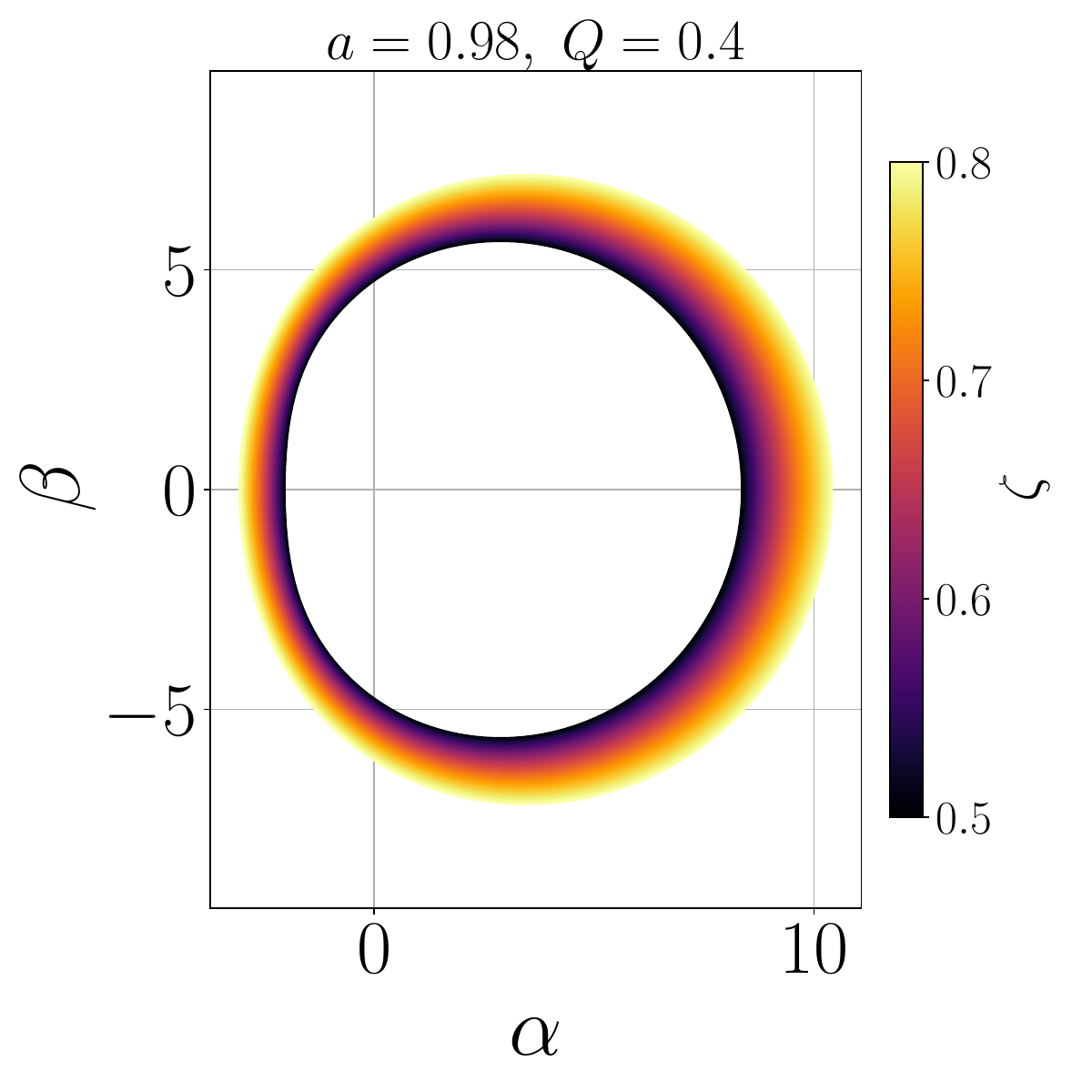}}
\centerline{\includegraphics[scale=0.35]{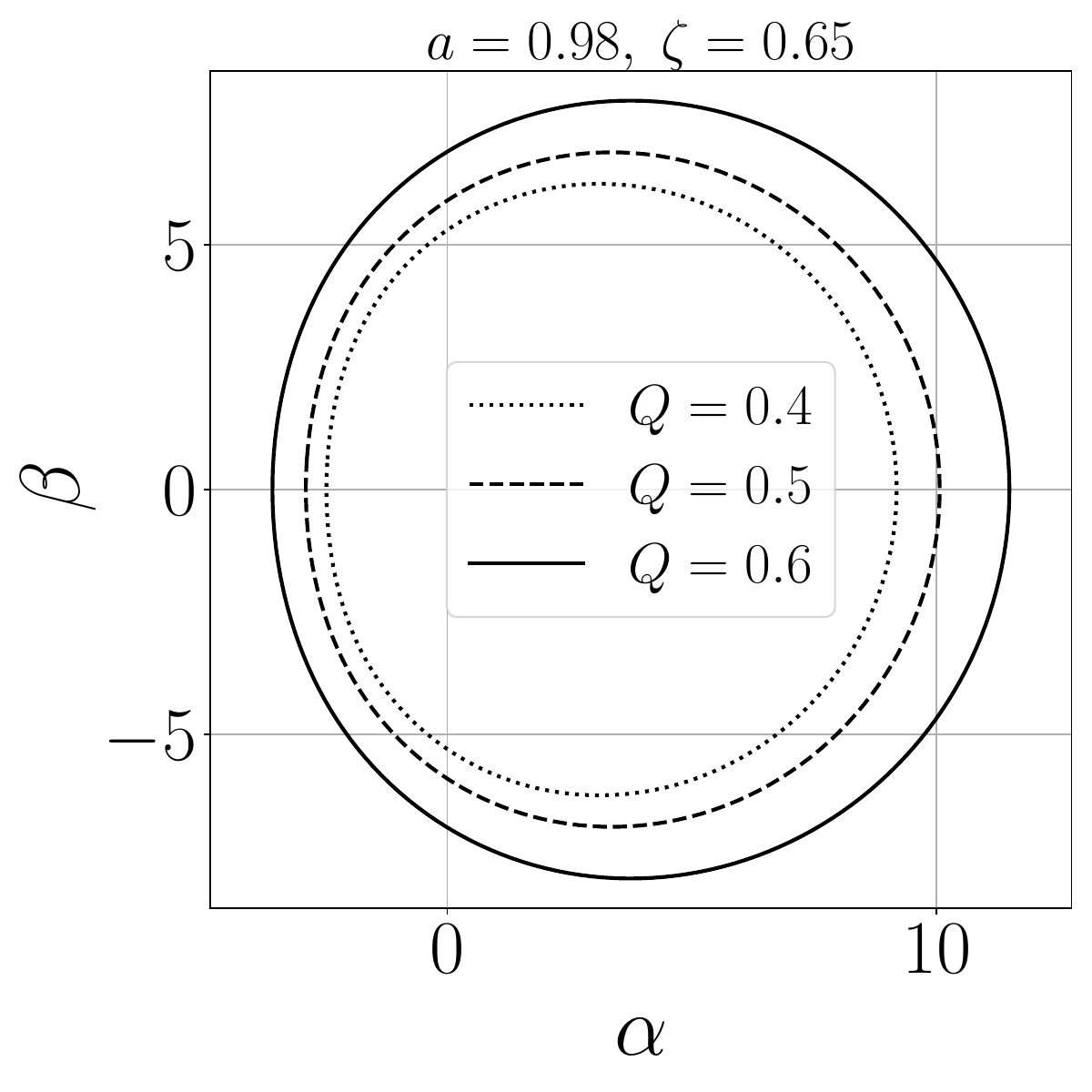}\includegraphics[scale=0.35]{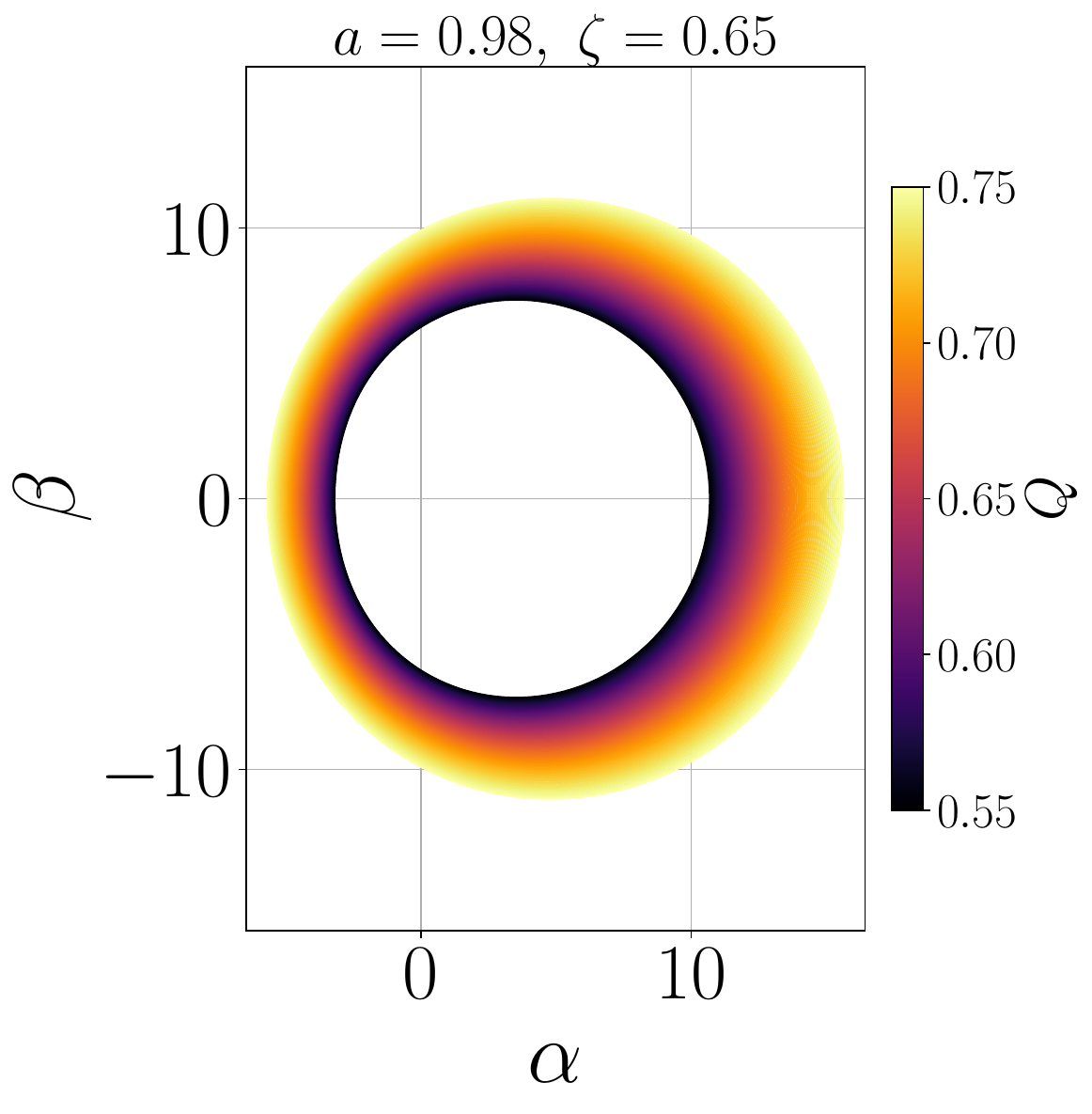}}
\caption{The shadow images are shown for the NLED BH with respect to the spin parameter, NLED parameter $\zeta$ and charge $Q$.}
\label{shadowsfig}
\end{figure*}

	The shadow images for the rotating NLED BH in the equatorial plane can be obtained by plotting Eq. \eqref{xy_shad2}. The images are shown in Fig. \ref{shadowsfig}. We have shown the shadow images for different choices of parameters. While keeping fixed values of $\zeta = 0.5$ and $Q = 0.4$, we plotted the shadows for three different values of spin parameters $a = 0.90, 0.95$ and $0.98$. We see that as the spin parameter increases, the shadow is gradually flattened towards the prograde end. Moreover, the size of the shadow decreases as $a$ increases. This is shown in the extreme left panel of Fig. \ref{shadowsfig}. With the increase of the spin parameter, the rotational energy of the BH generates stronger frame dragging, which in turn causes the surrounding spacetime to twist more and thus has a stronger effect on the nearby null orbits, thereby causing an asymmetrical bending of light that manifests as a flattening of the shadow on the nearer prograde end where the frame dragging effect is most intense. This stronger rotation further changes the effective photon sphere, shifting the structure of the co-rotating photon orbits inward, which leads to a reduction in the overall size of the apparent observable shadow.
	 Next, in the middle panel, the spin parameter $a$ is fixed to 0.98 and the charge $Q$ is set to 0.4. We observe, as the parameter $\zeta$ increases, the flattening of the nearer end of the shadow reduces. This suggests that the NLED modifications in the rotating framework are acting in a way that essentially counterbalances the frame-dragging effects caused by high spin values. This is one of the important results of this study. Physically, it implies that the NLED parameter modifies the gravito-electromagnetic interaction occurring within the system. Or in other words, the effective deviation from the standard Kerr geometry due to the introduction of NLED effects is reflected through the behaviour of the null rays. Finally, in the right panel, one can observe that the charge parameter $Q$ drastically affects the size of the BH shadow. As $Q$ increases, the size of the shadow increases, which counteracts the effective frame-dragging effect due to rotation. This suggests that the charge associated with the NLED BH plays an important role in shadow size determination from the observer's point of view.

\section{Conclusion}
\label{conc_sec}
In this paper, we have investigated the behaviour of null geodesics around an NLED BH through the backward ray tracing method. In the first part, we carried out a study of the null geodesics in the non‑rotating Schwarzschild-like BH spacetime with NLED corrections.  In the static, spherically symmetric situation, our backward ray tracing results show that the photon-sphere radius \(r_{ph}\) grows monotonically with both the NLED coupling parameter \(\zeta\) and the electric charge \(Q\) (Table~\ref{tab_null}). Increasing \(\zeta\) from 0.55 to 0.75 at constant \(Q=0.5\) increases the photon sphere radius \(r_{ph}\) outward through a small amount, and by a similar 0.20 increase in \(Q\) at constant \(\zeta=0.65\), one obtains an slightly larger increase in \(r_{ph}\). Nevertheless, both parameters play a similar role in the increase of the radius of the photon orbit.  The critical angular momentum \(L\) also necessarily increases in proportion, in accordance with an attenuated effective potential barrier. These results emphasize the leading contribution of the Coulombic term $Q$ over the NLED correction term $\zeta$ in redrawing the photon-sphere and hence indicate towards the possibility that BH shadow diameter precision measurements may simultaneously put constraints on the charge and beyond-Maxwell (or NLED) couplings. Our analysis of timelike geodesics around the NLED black hole shows that nonlinear electromagnetic corrections qualitatively change both binding and dynamical characteristics of massive particle orbits. Specifically, as the NLED parameter $\zeta$ or the charge $Q$ is increased, the effective potential well becomes increasingly deep, reducing the specific energy of stable circular orbits (and hence increasing binding energy) but pushing the ISCO radius inward. This deeper well also enhances the relativistic advance of perihelion -- measured in terms of geodesic precession frequency, as it reaches near the horizon and falls monotonically to zero as $r \to \infty$, with increasing effects of NLEDs giving a small but regular rise in precession at every radius. At distances far away, the velocity of orbit recovers to Newtonian $r^{-1/2}$ form, but relatively closer to the BH, it reaches its maximum value, moderately raised by increasing $\zeta$ and $Q$ values.

In the second part of the paper, by applying the Newman–Janis algorithm, we generalized our analysis to the rotating NLED black hole. The shape of the ergoregion shows that the BH has two separate horizons for any given pair of choices of \(\zeta\) and \(Q\), where the increase in the NLED correction $\zeta$ induces an increase in the gap between the two occurring horizons in the function \(\Delta\).  This suggests that, although the usual frame-dragging boundaries are mildly perturbed by the non-linearities, subtle changes in horizon geometry can still occur in strong-field environments.
Our shadow-image calculations in the equatorial plane (Fig.~\ref{shadowsfig}) also make clearer how the spin parameter $a$, NLED coupling term $\zeta$, and charge $Q$ work together to shape the visible shadows.  With increasing spin parameter \(a\), the shadow becomes increasingly flattened on the prograde side, and its total diameter is reduced as a result of increased frame dragging. Surprisingly, raising \(\zeta\) reverses this flattening, indicating that NLED corrections can help offset extremal spin effects to some extent by shifting the gravito‑electromagnetic coupling between null rays.  In the meantime, increasing \(Q\) expands the shadow radius, working against the shrinking of size induced by spin.

\bibliography{bibliography.bib}

\end{document}